\documentclass[a4paper,11pt]{article} 
\pdfoutput=1
\usepackage{jheppub} 
\usepackage[T1]{fontenc}
\usepackage[utf8]{inputenc} 
\usepackage[USenglish]{babel}
\usepackage{graphicx}  
\usepackage{float}
\usepackage{caption} 
\usepackage{subcaption}
\usepackage{multirow} 
\usepackage{amsmath,amsfonts,amssymb} 
\usepackage{tensor}
\usepackage{dsfont}
\usepackage{braket}
\usepackage{mathtools} 
\usepackage{siunitx}
\usepackage[dvipsnames]{xcolor}
\usepackage{booktabs}
\usepackage[capitalize,english]{cleveref}
\usepackage{xspace}
\usepackage{comment}
\usepackage{amsmath}

\DeclareMathAlphabet{\mathpzc}{OT1}{pzc}{m}{it}

\binoppenalty=\maxdimen
\relpenalty=\maxdimen 

\DeclareSIUnit\barn{b}

\DeclareMathOperator{\Tr}{Tr}
\newcommand{\Id}{\mathbb{I}}

\newcommand{\OO}{\mathcal{O}}

\newcommand*{\concurrence}{\ensuremath{\mathcal{C}}\xspace}
\newcommand*{\cUB}{\ensuremath{\mathcal{C}_\mathrm{UB}}\xspace}
\newcommand*{\cMB}{\ensuremath{\mathcal{C}_\mathrm{LB}}\xspace}
\newcommand*{\ie}{\emph{i.e.\@}\xspace}

\allowdisplaybreaks

\title{Probing new physics through entanglement in diboson production}

\preprint{IRMP-CP3-23-37} 
\author[a]{Rafael Aoude,}
\author[b]{Eric Madge,}
\author[a,c]{Fabio Maltoni,}
\author[d]{and Luca Mantani}
\affiliation[a]{Centre for Cosmology, Particle Physics and Phenomenology (CP3), \\ Universit\'e catholique de Louvain, 1348 Louvain-la-Neuve, Belgium}
\affiliation[b]{
Department of Particle Physics and Astrophysics,\\
Weizmann Institute of Science, Rehovot 7610001, Israel}
\affiliation[c]{Dipartimento di Fisica e Astronomia, Universit\`a di Bologna and INFN,\\ 
Sezione di Bologna, via Irnerio 46, 40126 Bologna, Italy}
\affiliation[d]{
DAMTP, University of Cambridge,\\ Wilberforce Road,
Cambridge, CB3 0WA, United Kingdom}
\emailAdd{rafael.aoude@uclouvain.be}
\emailAdd{eric.madge-pimentel@weizmann.ac.il}
\emailAdd{fabio.maltoni@uclouvain.be}
\emailAdd{luca.mantani@maths.cam.ac.uk}

\abstract{Pair production of heavy vector bosons is a key process at colliders: it allows to test our understanding of the Standard Model and to explore the existence of new physics 
through precision measurements of production rates and differential distributions. New physics effects can be subtle and often require observables specifically designed for their detection. In this study, we focus on quantum information observables that characterise the spin states of the final diboson system. We analyse concurrence bounds, purity, and Bell inequalities for a bipartite qutrit system representing two massive gauge bosons.  Our findings show that quantum spin observables can serve as complementary probes for heavy new physics as parametrised by higher dimensional operators in the Standard Model effective field theory. In particular, we find that these observables offer increased sensitivity to operators whose contributions do not interfere with the Standard Model amplitudes at the level of differential cross sections.}

\begin{document}

\maketitle

\newpage

\section{Introduction}
\label{sec:Intro}

Our current understanding of fundamental interactions of elementary particles is based on  relativistic quantum field theories (QFTs) that are built upon symmetry principles. Lorentz and Poincaré invariance, for example, not only dictate the possible particle content of the theory (in terms of group irreducible representations), but also strongly limit the possible form of their interactions.
Charge conservation and then gauge invariance, provide further constraints and non-trivial connections between interactions among different particles. Finally, the requirement of renormalisability further reduces the number of allowed interactions to a handful. The possible interactions in a renormalisable QFT is therefore very much constrained once the field content of the theory and its (gauge) symmetries are imposed. The most successful and famous example of a renormalisable QFT featuring a very limited number of interactions encoded in a simple Lagrangian, invariant with respect to $SU(3) \times SU(2) \times U(1)$ gauge symmetries, is the Standard Model (SM) of particle physics. 


\includecomment{Interactions among the fundamental constituents lead to non-trivial scattering amplitudes. The interaction of elementary particles is highly dependent on their quantum numbers -- charges, masses and spin -- and kinematical properties of a cross-section (or scattering amplitudes) give us insight on the underlying theory describing such events. The softness of a particle's momentum in an amplitude can unveil the underlying symmetries of the theory as well as its the high-energy limit, which can guide us on how to restore perturbative unitary.}

In addition to symmetries, other fundamental properties of QFTs, such as unitarity and positivity, have shown to provide very important constraints on the form of scattering amplitudes,  which can be obscure at the Lagrangian level. More recently, it has been suggested that being at the core of quantum mechanics, the entanglement properties of a system could be used to provide constraints on the underlying dynamics. For example, it was observed that the requirement of maximal entanglement puts constrains on the form of the  interactions in QED and the EW theory~\cite{Cervera-Lierta:2017tdt}, while in Refs.~\cite{Beane:2018oxh,Low:2021ufv} a very interesting relation between minimisation of entanglement and enhanced symmetries was observed for low energy QCD. These works focus on the entanglement of evolution dictated by the $S$-matrix, which acts a quantum logical gate, between the initial and the same final states. A different approach consists in studying the pattern of entanglement of a given final state in a generic scattering amplitude. The simplest example is to consider the spin degrees of freedom, described by a correlation matrix ($R$-matrix) which is pertubatively computable in QFT, and see how predictions change depending on the form of the interactions.  Although different, the above two approaches allow to explore the relation between symmetries and entanglement.

Recently, the authors of Ref.~\cite{Afik:2020onf} have pointed out that the quantum information properties of the spin states of top-anti-top quark pairs at proton colliders should be already accessible in current data.
Using two measures of entanglement, the concurrence and the Peres-Horodecki criterion, they identified  two phase space regions featuring maximal entanglement: at threshold and at high-$p_T$. This has triggered a series of studies on Standard Model $t\bar{t}$ production, that have further elaborated on the experimental detection strategy\,\cite{Fabbrichesi:2021npl,Severi:2021cnj,Afik:2022kwm,Aguilar-Saavedra:2022uye,Dong:2023xiw}. Moreover, other observables have been explored such as quantum steering and discord that allow the top-anti-top spin correlations~\cite{Afik:2022dgh}  to be organised in graded sets of quantum correlations characteristic of a two-qubit state:
\begin{align}
\text{Spin correlations} \supseteq  
\text{Discord} \supseteq  
\text{Entanglement}\supseteq  
\text{Steering}\supseteq  
\text{Bell Inequalities} \, .
\nonumber
\end{align}

In Ref.~\cite{Aoude:2022imd}, we have proposed to use quantum observables in $t\bar{t}$ to search for physics beyond the Standard Model~(BSM), \ie, to study the structure and properties of fundamental interactions at very high scales. Working in the SM Effective Field Theory~(SMEFT) framework, which allows to ``deform'' the SM in a consistent way (\ie compatible with the gauge symmetries and the particle content of the theory), we have calculated the new physics contributions proportional to the Wilson coefficients (with a linear and quadratic\footnote{Strictly speaking, by quadratic we here mean effects from the square of linear EFT amplitudes. In this work, we do not consider double dimension-six insertions or linear dimension eight, which are formally at the same order in the EFT expansion, \ie $\Lambda^{-4}$.} dependence), to  the concurrence fully analytically at the tree level. We have found that, in the SMEFT analysis of $t\bar t$ production, higher-dimensional operators reduce the entanglement generated by the SM.
The same conclusion was later corroborated also at next-to-leading order~(NLO) in $\alpha_S$, confirming the expectation that loop effects do not drastically change the leading-order~(LO) picture~\cite{Severi:2022qjy}. BSM effects in $t\bar{t}$ have also been explored in Ref.~\cite{Fabbrichesi:2022ovb}.

One of the reasons for interest in the top quark pair system is its simplicity: top quarks being fermionic ``bare'' spin-1/2 states, form bipartite qubit systems. Beyond $t\bar{t}$ production, other two-particle final states that feature spin correlations described by two qubits  have been proposed, from $\tau\tau$ to diphoton~\cite{Fabbrichesi:2022ovb,Altakach:2022ywa}. 

Qubits can model most of the SM particles -- fermions and massless bosons -- with the exception of the Higgs scalar and massive gauge bosons. Leaving out the Higgs boson which has no spin, one notes that $W^\pm$ and $Z$ bosons  being massive are characterised by three polarisations and therefore can be described by qutrits.  Barr et~al.~\cite{Barr:2021zcp,Barr:2022wyq,Ashby-Pickering:2022umy} initiated the quantum studies of final states involving massive vector bosons, first by introducing the qutrit formalism for entanglement at colliders and then exploring  Higgs boson decays and diboson production, the latter mainly studied using a numerical approach. The same processes have further been studied in Refs.~\cite{Fabbrichesi:2023cev, Aguilar-Saavedra:2022mpg, Aguilar-Saavedra:2022wam, Fabbrichesi:2023jep}. Recently, entanglement in diboson production was also studied in the context of vector-boson fusion~\cite{Morales:2023gow} and from decays of top pairs~\cite{Aguilar-Saavedra:2023hss}. The latter Ref. studies also for the first time the detection of entanglement between a $W$ boson and a top quark.

Quantifying entanglement through the concurrence~\concurrence~\cite{Hill:1997pfa,10.1063/1.2795840} is a challenging analytical task for qutrits, as   it involves an optimisation procedure, and closed analytic expressions can be only obtained in special cases or configurations. It turns out, however, that lower and upper bounds for the concurrence can be obtained in closed form,
\begin{align}
 \cMB \leq \concurrence(\rho) \leq \cUB,
\end{align}
where \cMB is the lower bound~\cite{PhysRevLett.98.140505} and \cUB the upper bound~\cite{Zhang_2008}, and that in some cases the bounds are so effective that they coincide with the actual values.

In this work we study the spin density matrix of diboson production, in the SM and in SMEFT, with the goal of understanding whether quantum observables may provide a better probe of new interactions than usual ``classical'' observables. 
This paper is organised as follows. In \cref{sec:Formalism}, we review the formalism used to study quantum information observables for spin-1 particles, which slightly differs from the one used for top quarks. This formalism is then applied to electroweak diboson production at present and future colliders. Their interactions in the SM and in the SMEFT are described in \cref{sec:interactions}. We proceed by studying perturbative unitarity and its relation with entanglement in \cref{sec:unitarity}. We finally present the results for diboson production at lepton and proton colliders in \cref{sec:Entanglement_colliders} and conclude in \cref{sec:Outro}. Details on the density matrix coefficients are given in the ancillary \texttt{Mathematica} notebook accompanying the {\tt arXiv} submission of this manuscript.

\section{Qutrit formalism}
\label{sec:Formalism}

The spin density matrix is a fundamental object in quantum mechanics as its knowledge completely characterises the quantum system. In this context, quantum tomography is the idea of conducting experiments with the objective of determining the density matrix of the system at study. This objective was at the heart of top quark pair spin studies in Refs.~\cite{Aoude:2022imd, Afik:2020onf, Severi:2021cnj,Afik:2022dgh,Afik:2022kwm, Aguilar-Saavedra:2022uye,Fabbrichesi:2021npl,Severi:2022qjy, Fabbrichesi:2022ovb}. In the aforementioned studies, several quantum information observables have been  analysed, from spin correlations to quantum entanglement and tests of Bell inequalities, both in the SM and exploring effects from heavy NP through SMEFT dimension six effects.

\subsection{Spin density matrix and quantum observables}

In the following we present the theoretical framework that is going to be used throughout the paper to describe the spin density matrix of a bipartite system consisting of two spin-1 massive particles, \ie, qutrits. The formalism employed closely follows that of  Ref.~\cite{Ashby-Pickering:2022umy}. While for spin-$1/2$ particles the measurement of the particle polarisation completely defines the spin density matrix, this is not the case anymore for spin-$1$ particles, as the polarisation vector determination is not enough to fully characterise the system. For a generic particle of spin $s$, the Hilbert space has dimension $d=2 s +1$. The density matrix $\rho$ is therefore a $d\times d$ matrix with $d^2 - 1$ free parameters (as $\Tr[\rho]=1$). This means that we can always decompose the generic one-particle spin density matrix with the generators of $SU(d)$, \ie, the generalised Gell-Mann matrices:
\begin{equation}
    \rho = \frac{1}{d} \Id +\sum_{i=1}^{d^2-1} a_i \lambda_i \, .
\end{equation}
As made explicit from the above decomposition, in order to fully characterise the quantum system one needs to determine the Bloch vector ${a_i}$. In the case of spin-$1/2$ particles, the Bloch vector has dimension $3$, the same of the spin operator $\vec{S}$. This means that the density matrix can be recast in terms of the spin operator spatial components and therefore its determination is in one-to-one correspondence with the quantum tomography of the system. 

The situation is more complicated for particles of higher spins. For instance, in the case of massive spin-$1$ particles, the Bloch vector has dimension $8$ and the measurement of the spin components of the particle are not sufficient anymore for the complete characterisation of the quantum state.
It is however possible to express the spin density matrix in terms of a spin matrix representation. In particular, given the spin-$1$ matrices
\begin{equation}
S_x=\frac{1}{\sqrt{2}}\left(\begin{array}{ccc}
0 & 1 & 0 \\
1 & 0 & 1 \\
0 & 1 & 0
\end{array}\right), \quad S_y=\frac{1}{\sqrt{2}}\left(\begin{array}{ccc}
0 & -\mathrm{i} & 0 \\
\mathrm{i} & 0 & -\mathrm{i} \\
0 & \mathrm{i} & 0
\end{array}\right), \quad S_z=\left(\begin{array}{ccc}
1 & 0 & 0 \\
0 & 0 & 0 \\
0 & 0 & -1
\end{array}\right) \, ,
\end{equation}
one can build a set of six operators
\begin{equation}
    S_{\{i j\}} \equiv S_i S_j+S_j S_i
\end{equation}
that together with the spin matrices allow us to decompose the spin density matrix in the form
\begin{equation}
    \rho=\frac{1}{3} \Id+\sum_{i=1}^3 \alpha_j S_i+\sum_{i, j=1}^3 \beta_{i j} S_{\{i j\}} \, .
\end{equation}
Note that in this formalism not all of the coefficients are free parameters since some of these operators are not linearly independent from the identity matrix, \ie
\begin{equation}
    S_{\{x x\}}+S_{\{y y\}}+S_{\{z z\}}=2\left(S_x^2+S_y^2+S_z^2\right)=4 \,\Id \, .
\end{equation}
In order for $\rho$ to have unit trace we therefore have to impose the constraint
\begin{equation}
    \sum_{i=1}^3 \beta_{ii} = 0 \, .
\end{equation}
The generalisation to a bipartite system in this formalism is straightforward. The spin density matrix for a pair of qutrits, in the Gell-Mann decomposition, is given by
\begin{equation}
\label{eq:FanoDecompositionNormalized}
    \rho=\frac{1}{9} \,\Id\otimes\Id + \frac{1}{3} \sum_{i=1}^{8} a_i \, \lambda_i \otimes  \Id + \frac{1}{3} \sum_{j=1}^{8} b_j\, \Id \otimes \lambda_j+\sum_{i=1}^{8} \sum_{j=1}^{8} c_{i j} \,\lambda_i \otimes \lambda_j \, .
\end{equation}
Similarly, one could express \cref{eq:FanoDecompositionNormalized} in terms of the spin matrix representation, but we find the Gell-Mann decomposition more straightforward and neat to use. 

The parameters $a_i$, $b_j$ and $c_{i j}$ (called Fano coefficients) determine the angular distributions of the decay products, and characterise the interactions governing the decay. This feature allows us to perform the quantum tomography of the system experimentally by measuring the angular distributions of the decays and reconstructing the coefficients. The aim of this work is to first explore the properties of the density matrix based on observables\footnote{By observables, we mean final states angular distributions after the gauge boson decays.} and the conditions for entanglement, deferring the tomography studies to a later stage. 

\subsection*{Concurrence}

We state that a system is entangled if the concurrence $\concurrence(\rho)$ is non zero. For bipartite qubit systems, this measure is easily evaluated by relating to the eigenvalues of the matrix $\omega = \sqrt{\sqrt{\tilde{\rho}}\rho \sqrt{\tilde{\rho}}}$ where $\tilde{\rho} = (\sigma_2\otimes \sigma_2)\rho^*(\sigma_2\otimes \sigma_2)$. However, for more complicated bipartite systems, such as the two qutrits explored in this work,  conditions for entanglement  cannot be calculated analytically~\cite{Uhlmann:1996mk,Horodecki:2009zz,Barr:2022wyq}.

For higher-dimensionality mixed states, the concurrence is obtained by the method of convex roof extension~\cite{Uhlmann:1996mk}. For a given decomposition of the $\rho$ matrix in pure states, \ie
\begin{align}
\label{eq:RhoDecompositionPureStates}
\rho = \sum_i p_i |\psi_i\rangle \langle \psi_i |\,, 
\qquad \sum_i p_i =1\,, 
\qquad p_i \geq 0\,,
\end{align}
the concurrence is defined as
\begin{align}
\label{eq:ConcurrenceInfimum}
\concurrence(\rho) = \text{inf}\left[ \sum_i p_i c(|\psi_i\rangle)\right] \,,
\end{align}
where the infimum is obtained over all possible ensembles $\{p_i,\psi_i\}$ for the decomposition in \cref{eq:RhoDecompositionPureStates}. The particular ensemble to which this infimum is reached is called optimal. The given concurrence is then the average of the optimal ensemble states concurrence.\footnote{It is clear form \cref{eq:ConcurrenceInfimum} that if we have a pure state, the concurrence will be calculable. This will be relevant later for $WZ$ production.}
The concurrence, however, cannot be calculated in a closed form for systems higher than $2 \times 2$. In these cases, one can rely on lower and upper bounds, \cMB and \cUB respectively, to quantify the entanglement.

A lower bound on the concurrence is given by~\cite{PhysRevLett.98.140505}
\begin{equation}
    (\concurrence(\rho))^2 \geq 2 \max\left(0, \Tr\left[\rho^2\right]-\Tr\left[\rho_A^2\right],\Tr\left[\rho^2\right] -\Tr\left[\rho_B^2\right]\right) \equiv \cMB^2 \, ,
\end{equation}
where $\rho_A = \Tr_B(\rho)$ and $\rho_B = \Tr_A(\rho)$ are the reduced density matrices, obtained by tracing out one of the subsystems. The function $\cMB(\rho)$ is a marker that is telling us that in case of positive values, we have an entangled state. However, in the scenario of negative values, the test is inconclusive. In particular, for the qutrit pair, we have
\begin{equation}
\label{eq:cmb}
    \cMB^2=-\frac{4}{9}+\max\left(-\frac{8}{3} \sum_{i=1}^8 a_i^2+\frac{4}{3} \sum_{j=1}^8 b_j^2, \frac{4}{3} \sum_{i=1}^8 a_i^2-\frac{8}{3} \sum_{j=1}^8 b_j^2\right)+8 \sum_{i, j=1}^8 c_{i j}^2 \, .
\end{equation}
One can also obtain an upper bound for the concurrence, given in Ref.~\cite{PhysRevA.78.042308} and recently explored also in Ref.~\cite{Fabbrichesi:2023cev},
\begin{equation}
(\concurrence(\rho))^2 \leq 2 \, \text{min} \, \left(1- \text{Tr}[\rho_A^2], 1- \text{Tr}[\rho_B^2] \right) \equiv \cUB^2 \, ,
\end{equation}
which in terms of the Fano coefficients reads
\begin{equation}
\label{eq:cub}
 \cUB^2 = \frac{4}{3}-4 \, \text{min} \left(\sum_{i=1}^8 a_i^2, \sum_{j=1}^8 b_j^2 \right) \,.
\end{equation}

For a qutrit pair, the maximum value of the concurrence is obtained for a totally symmetric and entangled pure state,
\begin{equation}
    \left|\Psi_{+}\right\rangle=\frac{1}{\sqrt{3}} \sum_{i=1}^3|i\rangle \otimes|i\rangle \,,
\end{equation}
with $\concurrence(\rho)=2/\sqrt{3}$. This is different from a qubit pair, in which all entangled pure states have concurrence $\concurrence(\rho)=1$.

\subsection*{Purity}

When exploring the density matrices to assess entanglement, it is useful to know if the state is pure or mixed. This is quantified by the purity~$P$ given by
\begin{align}
    P(\rho) \equiv \text{tr}[\rho^2]\,,
\end{align}
which is one in the case of pure states and bounded to the lower value of $1/d$ for qudits. Given the Fano decomposition in \cref{eq:FanoDecompositionNormalized}, this means
\begin{align}
P(\rho)= \frac{1}{9} 
+ \frac{2}{3}\sum_{i=1}^8 (a_i^2+b_i^2) 
+ 4 \sum_{i,j=1}^8 c_{ij}^2 \,.
\end{align}

\subsection*{Bell inequalities}
Another interesting aspect of quantum system is the possibility of violating Bell inequalities~\cite{Bell:1964kc}. This allows to distinguish classical local realist theories from quantum mechanical ones. In particular, for a pair of qubits, the  Clauser-Horne-Shimony-Holt (CHSH)~\cite{Clauser:1969ny} inequality holds
\begin{equation}
    \mathcal{I}_2=E(a, b)-E\left(a, b^{\prime}\right)+E\left(a^{\prime}, b\right)+E\left(a^{\prime}, b^{\prime}\right) \leq 2 \, .
\end{equation}
Quantum mechanics allows  $\mathcal{I}_2$ to have values higher than two. 

Analogously, one can define an observable for pairs of qutrits, the Collins-Gisin-Linden-Massar-Popescu (CGLMP) inequality~\cite{PhysRevA.65.032118, PhysRevLett.88.040404}. By defining the quantum operator~\cite{PhysRevLett.68.3259, Acin:2002zz}
\begin{equation}
    \mathcal{B}=-\frac{2}{\sqrt{3}}\left(S_x \otimes S_x+S_y \otimes S_y\right)+\lambda_4 \otimes \lambda_4+\lambda_5 \otimes \lambda_5 \, ,
\end{equation}
one finds the CGLMP inequality
\begin{equation}
    \mathcal{I}_3 = \Tr[\rho \mathcal{B}] \leq 2 \, .
\end{equation}
The above equation is valid in the $x-y$ plane, but it can be generalised to any direction in the 3-dimensional space and arbitrary bases in spin space. The generalised condition for violation of the Bell inequalities becomes
\begin{equation}
    \left\langle \mathcal{B}\right\rangle_\mathrm{max }=\max_{U,V}\left(\Tr\left(\rho \, (U^{\dagger} \otimes V^{\dagger}) \, \mathcal{B} \, (U \otimes V)\right)\right) \geq 2 \, ,
\end{equation}
where $U,V \in U(3)$ are unitary matrices.

\subsection{EW boson production at colliders}

We now turn our attention to electroweak production of spin-$1$ particles at colliders, \ie $W$ and $Z$ bosons. Following the approach of Ref.~\cite{Aoude:2022imd}, we define the $R$-matrix from the matrix element amplitude:
\begin{align}
        R_{\alpha_1 \alpha_2, \beta_1 \beta_2}^I &\equiv \frac{1}{N_a N_b} \sum_{\substack{\text { colors } \\ \text { a,b spins }}} \mathcal{M}_{\alpha_2 \beta_2}^* \mathcal{M}_{\alpha_1 \beta_1} \\
        \text { with } \quad \mathcal{M}_{\alpha \beta} &\equiv\left\langle V\left(k_1, \alpha\right) \bar{V}\left(k_2, \beta\right)|\mathcal{T}| a\left(p_1\right) b\left(p_2\right)\right\rangle \nonumber \, ,
\end{align}
where $I= ab=\bar{q}q', \bar{e}e$ are the possible initial states in proton and lepton colliders at LO with $N_{a,b}$ degrees of freedom, and $V=W^\pm,\,Z$ are the respective vector bosons. For all these diboson amplitudes, we can factor out the polarisation vectors, which carry the spin dependence of the $R$-matrix as
\begin{align}
\label{eq:Amplitude_Polvec}
\mathcal{M}_{\alpha\beta} = \mathcal{M}_{\mu\nu} \,\,\varepsilon^{\dagger\mu}_\alpha(k_1)\varepsilon^{\dagger\nu}_\beta(k_2)
\end{align}
where both polarisation tensors act as a map between the Lorentz tensor structures in $\mathcal{M}_{\mu\nu}$ and the spin-space labelled by the index $\{\alpha,\beta\}$ forming the $9 \times 9$ qutrit matrix.
The $R$-matrix is in direct relation with the spin density matrix, \ie
\begin{equation}
    R = \tilde{A} \, \Id\otimes\Id + \sum_{i=1}^{8} \tilde{a}_i \,\lambda_i \otimes  \Id + \sum_{j=1}^{8} \tilde{b}_j\, \Id \otimes \lambda_j+\sum_{i=1}^{8} \sum_{j=1}^{8} \tilde{c}_{i j}\, \lambda_i \otimes \lambda_j \, ,
\end{equation}
with the coefficient $\tilde{A}$ encoding information on the differential cross section
\begin{equation}
    \frac{\mathrm{d} \sigma}{\mathrm{d} \Omega}=\frac{9\beta}{64 \pi^2\hat{s}} \tilde{A}(\hat{s}, \boldsymbol{k}) \, ,
\end{equation}
where $\boldsymbol{k}$ is the direction of the $V$ boson, $\hat{s}$ the invariant mass of the pair and $\beta=\sqrt{1-4 \frac{m_V^2}{\hat{s}}}$ the velocity of the $V$ boson in the centre of mass frame.
Each coefficient can be obtained by tracing with the element of decomposition, e.g. $\tilde{a}_i = \text{tr}[R\,\lambda_i \otimes  \Id]$
and similarly for $\tilde{b}_j$ and~$\tilde{c}_{ij}$. 

If we consider production at proton colliders, such as the LHC, the total $R$-matrix is  given by a weighted sum of the various different partonic channels, \ie
\begin{equation}
    R(\hat{s}, \boldsymbol{k})=\sum_I L^I(\hat{s}) R^I(\hat{s}, \boldsymbol{k}) \, ,
\end{equation}
with $L_I$ the luminosity functions~\cite{Bernreuther:1997gs}. The relevant channels for diboson production at a proton collider are the quark annihilation ones. Note that, given that the initial state particles are not identical, both $q\bar{q}$ and $\bar{q}q$ channels are to be summed over. 
This can be also taken into account by a symmetrisation over the polar angle, since it can be shown that $R^{\bar{q}q}(\hat{s}, \theta)=R^{q\bar{q}}(\hat{s}, \theta + \pi)$, \ie
\begin{equation}
    R(\hat{s}, \theta)=\sum_q L^{q\bar{q}}(\hat{s})( R^{q\bar{q}}(\hat{s}, \theta) + R^{q\bar{q}}(\hat{s}, \theta + \pi)) \, ,
    \label{eq:Rmat_pp}
\end{equation}
where we made explicit that there is no dependence on the azimuthal angle $\phi$ of the vector $\boldsymbol{k}$, given the cylindrical symmetry of the problem. It is clear from the expression that the $R$-matrix of a proton collider is by definition symmetric around $\theta=\pi/2$.
The $R$-matrix is then related to the spin density matrix simply by an overall normalisation factor, \ie $\rho= R/\Tr(R)$. In particular we obtain the decomposition of the spin density matrix in terms of the $R$-matrix Fano coefficients
\begin{align}
    a_i =\frac{\tilde{a}_i}{3 \tilde{A}} \, , \qquad
    b_i =\frac{\tilde{b}_i}{3 \tilde{A}} \, , \qquad
    c_{ij} =\frac{\tilde{c}_{ij}}{9 \tilde{A}} \, .
\end{align}
In terms of these coefficients, the purity condition $P(\rho)=1$ reads
\begin{align}
36\tilde{A}^2 =
3\sum_{i=1}^8 (\tilde{a}_i^2+\tilde{b}_i^2) 
+ 2\sum_{i,j=1}^8 \tilde{c}_{i,j}^2 \,.
\end{align}

We conclude this section by commenting on the possible effects of higher-order corrections which in general will change the $R$-matrix and the expected entanglement. In specific cases, such as $ZZ$ and $W^+W^-$ final states, IR/UV finite loop-induced processes could also contribute, e.g., $gg\to ZZ/W^+W^-$. Even though suppressed, at the LHC gluon fusion production provides interesting information on Higgs properties. The framework presented here could be directly employed to perform a dedicated study. This  is left for future investigations. More in general, for QCD or QED corrections, real and virtual contributions need to be considered together. In case of inclusive predictions, the framework presented here can be can be straightforwardly applied by simply tracing out (including integration over the phase space) the unobserved degrees of freedom. Naively, one can expect some degree of decoherence which should lower the entanglement compared to the leading-order $R$-matrix. On the other hand, final states characterised by resolvable emissions could be analysed on their own as three-body final states, where all vector bosons are measured and the $R$-matrix has dimensionality higher than $9\times 9$. In this case the an extension of the framework presented here would be needed.

\section{Diboson interactions}
\label{sec:interactions}

In this section, we discuss the relevant interactions for diboson production at colliders. In particular, we consider the case of the SM as well as its extension in the SMEFT, where higher dimensional operators will lead to the introduction of SM parameter shifts and new Lorentz structures. The objective is to present the structure of the EW couplings dictated by the SM symmetries and how heavy new physics could potentially alter it.

\subsection{SM couplings}

\begin{figure}[t!]
\begin{minipage}{\textwidth}
 \centering
 \includegraphics{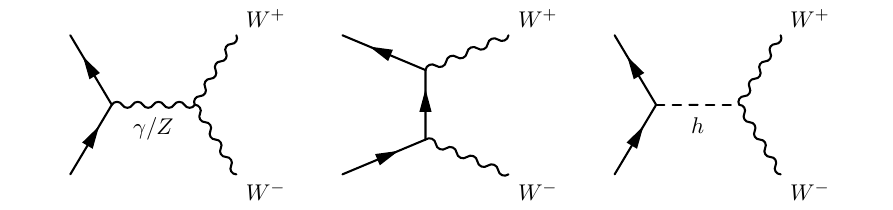}
 \caption{$W^{+}W^{-}$ production in the SM.}
 \label{fig:diagrams_WW}
\end{minipage}

\vspace{5mm}
\begin{minipage}{.49\textwidth}
 \centering
 \includegraphics{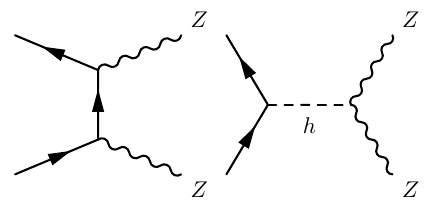}
 \caption{$ZZ$ production in the SM.}
 \label{fig:diagrams_ZZ}
\end{minipage}
\hfill
\begin{minipage}{.49\textwidth}
 \centering
 \includegraphics{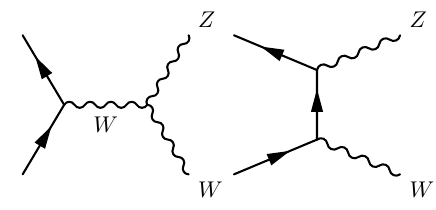}
 \caption{$WZ$ production in the SM.}
 \label{fig:diagrams_WZ}
\end{minipage}
\end{figure}

As we are interested in both lepton and hadron colliders, we now discuss the couplings that EW bosons have with quarks and with electrons. In particular, we consider the processes $e^+ e^- \to W^+ W^-$ and $e^+ e^- \to Z Z$, for a lepton collider as well as $p p \to W^+ W^-$, $p p \to Z Z$, and $p p \to W^+ Z$ for an hadron collider, respectively. In  the latter case, the relevant partonic channels at LO are given by $u\bar{u}$ and $d\bar{d}$ for the neutral final states, while for $W^+ Z$ it is $u \bar{d}$. We work in the 5-flavour scheme, so $u$ comprises both up and charm quark, while $d$ includes down, strange and bottom quarks. We choose to discuss $W^+ Z$, in analogy with Ref.~\cite{Fabbrichesi:2023cev}, as it is the dominant channel in a proton collider, but kinematic features are similar for $W^- Z$. 

In \cref{fig:diagrams_WW,fig:diagrams_ZZ,fig:diagrams_WZ} we show the topology of interest in terms of Feynman diagrams for all the processes considered. The properties of each process depend on the values and form of couplings, which, in the SM, are completely determined by gauge symmetry and the EWSB pattern. Specifically, the relevant couplings of the fermions are the coupling to the $Z$ boson, the coupling to the $W$ boson and the coupling to the $h$ boson. The latter is proportional to the mass of the fermions, so it will be highly suppressed and substantially irrelevant for the phenomenology at hadron or $e^+e^-$ colliders. We will however discuss their role in the high energy limit in \cref{sec:unitarity}.
Moreover, the processes are sensitive to the triple gauge coupling~(TGC) between the $W$ and the $Z$ boson, as well as the coupling of the $W$ with a photon. The coupling of EW bosons to the $h$ boson is always accompanied by the Yukawa coupling of the fermions, therefore our sensitivity to that is negligible\footnote{However it could be interesting at a future Higgs factory at $\sqrt{s}=m_h$}. The couplings of the fermions to the EW bosons read
\begin{align}
    \mathcal{L}_{\bar{f}fZ} \propto \left(g_V^Z\right)-\left(g_A^Z\right) \gamma_5 \,,
    \qquad\text{with}\qquad
    g_V^Z = \frac{T_3}{2} - Q \sin^2{\theta_W} \,, \quad
    g_A^Z = \frac{T_3}{2} ,\label{eq:Z_coupl}
\end{align}
for the $Z$ boson and
\begin{align}
    \mathcal{L}_{\bar{f}fW} &\propto g_W \left(1 - \gamma_5\right) \, ,\label{eq:W_coupl}
\end{align}
for the $W$ boson, with $T_3=-1/2$ for down-type quarks and leptons, $T_3=1/2$ for up-type quarks and the electric charge $Q=-1, 2/3, -1/3$ for $e$, $u$ and $d$ respectively. In the SM, $g_V^Z \approx -0.027, 0.1, -0.17$ and $g_A^Z \approx -0.25, 0.25, -0.25$ for $e$, $u$ and $d$ respectively and $g_W=1/2$.
Finally, the TGCs are specified by the gauge symmetries of the SM through the gauge Lagrangian
\begin{equation}
    \mathcal{L}_{\text {gauge }}=-\frac{1}{4} W_{\mu \nu}^a W^{a \mu \nu}-\frac{1}{4} B_{\mu \nu} B^{\mu \nu} \, .
\end{equation}
In the SM, the couplings are given by
\begin{align}
    g_{WW\gamma} = e \, , \qquad
    g_{WWZ} = e \cot{\theta_W} \, , 
\end{align}
where the electric charge $e$ and the trigonometric functions of the Weinberg angle can be determined in terms of the chosen EW input parameters.

\subsection{Modified interactions: SMEFT framework}
\begin{table}[t!]
  \begin{center}
    \renewcommand{\arraystretch}{1.4}
    {\small
    \begin{tabular}{cccc}
      \toprule
      Operator $\qquad$ & Coefficient & Definition & 95\,\% CL bounds\\
      \hline\bottomrule
    \multicolumn{4}{c}{two-fermion operators}\\
    \toprule
    $\OO_{\varphi u}$ & $c_{\varphi u}$ & $i\big(\varphi^\dagger \overset{\leftrightarrow}{D}_\mu\varphi\big)\big(\bar{u}\gamma^\mu u\big)$ & $\left[-0.17,0.14\right]$ \\\hline
    $\OO_{\varphi d}$ & $c_{\varphi d}$ & $i\big(\varphi^\dagger \overset{\leftrightarrow}{D}_\mu\varphi\big)\big(\bar{d}\gamma^\mu d\big)$ & $\left[-0.07,0.09\right]$ \\\hline
    $\OO_{\varphi q}^{(1)}$ & $c_{\varphi q}^{(1)}$ & $i\big(\varphi^\dagger \overset{\leftrightarrow}{D}_\mu\,\varphi\big)
 \big(\bar{q}\,\gamma^\mu\,q\big)$ & $\left[-0.06,0.22\right]$ \\\hline
    $\OO_{\varphi q}^{(3)}$ & $c_{\varphi q}^{(3)}$  & $i\big(\varphi^\dagger\overset{\leftrightarrow}{D}_\mu\,\tau_{ I}\varphi\big)
 \big(\bar{q}\,\gamma^\mu\,\tau^{ I}q\big)$ & $\left[-0.21,0.05\right]$ \\ \hline
    $\OO_{\varphi e}$ & $c_{\varphi e}$ & $i\big(\varphi^\dagger \overset{\leftrightarrow}{D}_\mu\varphi\big)\big(\bar{e}\gamma^\mu e\big)$ & $\left[-0.21,0.26\right]$ \\\hline
    $\OO_{\varphi l}^{(1)}$ & $c_{\varphi l}^{(1)}$ & $i\big(\varphi^\dagger \overset{\leftrightarrow}{D}_\mu\varphi\big)\big(\bar{l}\gamma^\mu l\big)$ & $\left[-0.11,0.13\right]$\\\hline
    $\OO_{\varphi l}^{(3)}$ & $c_{\varphi l}^{(3)}$ & $i\big(\varphi^\dagger \overset{\leftrightarrow}{D}_\mu \tau_I\varphi\big)\big(\bar{l}\gamma^\mu \tau^I l\big)$ & $\left[-0.21,0.05\right]$\\\hline
 \bottomrule
 \multicolumn{4}{c}{bosonic operators}\\
 \toprule
$\OO_W$ &  $c_{W}$ & $\varepsilon_{IJK} W^I_{\mu\nu}W^{J,\nu\rho} W^{K,\mu}_\rho$,  & $\left[-0.18,0.22\right]$ \\\hline
$\OO_{\varphi W}$ &   $c_{\varphi W}$ & $\bigg( \varphi^\dagger\varphi - \frac{v^2}{2}\bigg) W_I^{\mu\nu} W_{\mu\nu}^I$ & $\left[-0.15,0.30\right]$ \\\hline
$\OO_{\varphi B}$ &   $c_{\varphi B}$ & $\bigg( \varphi^\dagger\varphi - \frac{v^2}{2}\bigg) B_{\mu\nu} B^{\mu\nu}$ & $\left[-0.11,0.11\right]$\\\hline
$\OO_{\varphi WB}$ &   $c_{\varphi W B}$ & $(\varphi^\dagger \tau_I \varphi) B^{\mu\nu} W^I_{\mu\nu}$ & $\left[-0.17,0.27\right]$ \\\hline
$\OO_{\varphi D}$ &   $c_{\varphi D}$ & $(\varphi^\dagger D^\mu \varphi)^\dagger (\varphi^\dagger D_\mu \varphi)$ & $\left[-0.52,0.43\right]$ \\\hline
  \bottomrule
  \multicolumn{4}{c}{four-fermion operator}\\
  \toprule
  $\OO_{ll}$ &  $c_{ll}$ & $\big(\bar{l}\gamma_\mu l\big)\big(\bar{l}\gamma^\mu l\big)$ & $\left[-0.16,0.02\right]$   \\
  \hline\bottomrule
\end{tabular}
}
\end{center}
  \caption{Definition of the dimension-six SMEFT operators relevant for this analysis.
    The bounds assume a scale of $\Lambda=\SI{1}{\TeV}$ and are taken from the global fit of Ref.~\cite{Ethier:2021bye} at order $\mathcal{O}(\Lambda^{-4})$. The bound on $c_{ll}$ comes from EWPO fits and here we quote the result from Ref.~\cite{Brivio_2017}.
    \label{tab:ops}}
\end{table}

We now discuss the effects of heavy new physics to the production of diboson at colliders. We do so within the framework of the SMEFT where the SM gauge symmetries are all preserved and the Higgs mechanism is realised linearly. The SMEFT is an extension of the SM in which higher order operators modify the SM interactions, characterised by a Lagrangian of the kind
\begin{equation}
    \mathcal{L}_{\mathrm{SMEFT}}=\mathcal{L}_{\mathrm{SM}}+\sum_{n=1}^N \frac{c_n}{\Lambda^2} \mathcal{O}_n + \mathcal{O}\left(\frac{1}{\Lambda^4}\right) \, ,
\end{equation}
where $\mathcal{O}_n$ indicates a higher dimensional operator, $c_n$ is the associated Wilson coefficient, a free parameter that in a bottom-up approach needs to be determined experimentally, and $N$ is the number of operators at this order in the expansion. 

The operators are suppressed by the scale of new physics $\Lambda$, assumed to be at least at the order of a TeV. This scale is what allows us to truncate the EFT series by means of its power counting.
For this analysis, we focus on the SMEFT dim-6 operators~\cite{Grzadkowski:2010es,Aguilar-Saavedra:2018ksv} at order $1/\Lambda^2$ and neglect higher order corrections. Note that we can only be sensitive to the ratio $c_n/\Lambda^2$ and therefore we will absorb $\Lambda^2$ into the definition of the Wilson coefficient for the rest of the paper, \ie
\begin{equation}
    \frac{c_n}{\Lambda^2} \to c_n \, .
\end{equation}
For the sake of simplicity, we assume flavour universality for the few operators that could, in principle, have a more involved structure.
Results will be presented in the $m_W$ input parameter scheme, where the  experimentally determined EW parameters of choice are $\{m_W, m_Z, m_h, G_F\}$.
In \cref{tab:ops} the relevant CP-even operators for leading-order diboson production, both in lepton and proton colliders are shown.
We note that when considering a massless initial state, $\OO_{\varphi W}$ and $\OO_{\varphi B}$ which modify the coupling of the EW bosons to the Higgs, do not enter the processes because of the aforementioned pairing with the Yukawa couplings. We will therefore not discuss further these two operators, but we listed them for completeness.

The operators act in a multitude of ways, leading to different phenomenological consequences.
In particular, some operators act by shifting the SM value of the coupling of the fermions to the EW bosons. Specifically we have
\begin{align}
    \delta g_V^Z(e) = &\left(\sin^2{\theta_W}-\frac{1}{4}\right) \delta g_Z - \delta s^2_\theta -\frac{c_{\varphi l}^{(3)}+c_{\varphi e}+c_{\varphi l}^{(1)}}{4 \sqrt{2} G_F} \, , \notag\\
   \delta g_A^Z(e) = &\frac{\delta g_Z}{4}-\frac{c_{\varphi l}^{(3)}-c_{\varphi e}+c_{\varphi l}^{(1)}}{4\sqrt{2} G_F} \, , \notag\\
   \delta g_V^Z(u) = &\left(\frac{1}{4}-\frac{2 \sin^2{\theta_W}}{3}\right) \delta g_Z +\frac{2}{3} \delta s^2_\theta  -\frac{c_{\varphi q}^{(-)}+c_{\varphi u}}{4 \sqrt{2} G_F} \, , \\
   \delta g_A^Z(u) = &\frac{\delta g_Z}{4}-\frac{c_{\varphi q}^{(-)}-c_{\varphi u}}{4 \sqrt{2} G_F} \, ,\notag\\
   \delta g_V^Z(d)= &\left(\frac{\sin^2{\theta_W}}{3}-\frac{1}{4}\right) \delta g_Z -\frac{1}{3} \delta s^2_\theta -\frac{2 c_{\varphi q}^{(3)}+c_{\varphi d}+c_{\varphi q}^{(-)}}{4 \sqrt{2} G_F} \, , \notag\\
   \delta g_A^Z(d)= & \frac{\delta g_Z}{4}-\frac{2 c_{\varphi q}^{(3)}-c_{\varphi d}+c_{\varphi q}^{(-)}}{4
   \sqrt{2} G_F} \, ,\notag
\end{align}
where we defined $c_{\varphi q}^{(-)} = c_{\varphi q}^{(1)} - c_{\varphi q}^{(3)}$, which is the combination of Wilson coefficients that modifies the coupling of up-type quarks to the $Z$ boson. The variations of the couplings are written as function of the quantities
\begin{align}\begin{aligned}
    \delta g_Z &= -\frac{4 c_{\varphi l}^{(3)}-2 c_{ll}+c_{\varphi D}}{4 \sqrt{2}G_F} \, , \\
    \delta s^2_\theta &= \frac{c_{\varphi D} \, m_W^2}{2 \sqrt{2} G_F \, m_Z^2}+\frac{c_{\varphi W B} \, m_W \sqrt{1-\frac{m_W^2}{m_Z^2}}}{\sqrt{2} G_F \, m_Z} \, ,
\end{aligned}\end{align}
which are SMEFT induced universal shifts specific to the EW input parameter scheme, see Ref.~\cite{Brivio_2017} for more details.
For the coupling to the $W$ boson, we have
\begin{align}\begin{aligned}
    \delta g_W(e) &= \frac{c_{\varphi l}^{(3)}}{2 \sqrt{2} G_F} - \frac{\delta G_F}{2 \sqrt{2}} \, , \\
    \delta g_W(q) &= \frac{c_{\varphi q}^{(3)}}{2 \sqrt{2} G_F} - \frac{\delta G_F}{2 \sqrt{2}} \, ,
\end{aligned}\end{align}
with the fractional shift to the Fermi constant originating from the muon decay measurement given by
\begin{equation}
    \delta G_F = \frac{2 c_{\varphi l}^{(3)} - c_{ll}}{2 G_F} \, .
\end{equation}
Finally, the operators can also alter the TGCs. The modified Lagrangian reads
\begin{equation}
    \frac{\mathcal{L}_{W W V}}{-i g_{W W V}}=g_1^V\left(W_{\mu \nu}^{+} W^{-\mu} V^\nu-W_\mu^{+} V_\nu W^{-\mu \nu}\right)+\kappa_V W_\mu^{+} W_\nu^{-} V^{\mu \nu}+\frac{i \delta\lambda_V}{m_W^2} V^{\mu \nu} W_\nu^{+\rho} W_{\rho \mu}^{-} \, ,
\end{equation}
where $V = Z, \gamma$ and we have defined $V_{\mu \nu}=\partial_\mu V_\nu-\partial_\nu V_\mu$ and $W_{\mu \nu}^{ \pm}=\partial_\mu W_\nu^{ \pm}-\partial_\nu W_\mu^{ \pm}$. The dimension-6 SMEFT operators introduce a dependence on the couplings $g_1^V = 1 + \delta g_1^V$ and $\kappa_V = 1 + \delta \kappa_V$, \ie
\begin{align}\begin{aligned}
\delta g_1^\gamma & =\frac{1}{4 \sqrt{2} G_F}\left(c_{\varphi D} \frac{m_W^2}{m_W^2-m_Z^2}-4 c_{\varphi l}^{(3)}+2 c_{ll}-c_{\varphi W B} \frac{4 m_W}{\sqrt{m_Z^2-m_W^2}}\right) \, , \\
\delta g_1^Z & =\frac{1}{4 \sqrt{2} G_F}\left(c_{\varphi D}-4 c_{\varphi l}^{(3)}+2 c_{ll}+4 \frac{m_Z}{m_W} \sqrt{1-\frac{m_W^2}{m_Z^2}} c_{\varphi W B}\right) \, , \\
\delta \kappa_\gamma & =\frac{1}{4 \sqrt{2} G_F}\left(c_{\varphi D} \frac{m_W^2}{m_W^2-m_Z^2}-4 c_{\varphi l}^{(3)}+2 c_{ll}\right) \, , \\
\delta \kappa_Z & =\frac{1}{4 \sqrt{2} G_F}\left(c_{\varphi D}-4 c_{\varphi l}^{(3)}+2 c_{ll}\right) \, .
\end{aligned}\end{align}
Note that while $c_{\varphi l}^{(3)}$, $c_{ll}$ and $c_{\varphi D}$ universally shift the TGC of the SM, $c_{\varphi W B}$ does not contribute to $\kappa_V$ but only to $g_1^V$, changing the symmetrical structure of the SM interactions.
Only one operator in \cref{tab:ops} leads to a new Lorentz structure by generating a term proportional to $\delta \lambda_V$, with a dependence given by 
\begin{align}\begin{aligned}
    \delta \lambda_\gamma &= - 6 \sin{\theta_W} \frac{m_W^2}{g_{W W \gamma}} c_W \, ,\\
    \delta \lambda_Z &= - 6 \cos{\theta_W} \frac{m_W^2}{g_{W W Z}} c_W \, .
\end{aligned}\end{align}
$\OO_{W}$ is therefore of particular interest, since it modifies the interactions among EW bosons in a way that could potentially induce different helicity structures. This is of relevance for this study, given that the density matrix of the diboson system could markedly change if the EW bosons are produced in configurations not present in the SM.

\section{Perturbative unitarity and entanglement}
\label{sec:unitarity}

Perturbative unitarity and the role of the Higgs boson in the SM can be effectively studied by considering multi-boson longitudinal scattering amplitudes and their cross-section~\cite{Romao:2016ien}.
Conservation of probability in QFT requires the unitarity of the $S$-matrix, which imposes bounds on the energy dependence of the corresponding terms in the perturbative expansion.
The scattering amplitudes and cross-sections involving weak bosons violate the perturbative bound if gauge invariance is not respected or Higgs-mediated interactions are not included. How these violations appear and then cancel among different contributions in usual observables has been studied extensively. 
However, to the best of our knowledge, the perturbative unitarity constraints on entanglement and quantum information observables  has not been explored. Before proceeding, we note that for this study and at variance with what done in the previous sections, here  we keep the fermion masses non-zero, to allow non-vanishing  Higgs couplings and investigate their role too. 

It is possible to express 2$ \to$2 scattering amplitudes in powers of the normalised energy $\sqrt{s}/(2m_V)$: 
\begin{align}
\label{eq:Amplitude_HE_expansion}
    \mathcal{M} = \mathcal{M}^{(2)} \left( \frac{\sqrt{s}}{2m_V} \right)^2 + 
   \mathcal{M}^{(1)} \left( \frac{\sqrt{s}}{2m_V} \right) +
   \mathcal{M}^{(0)} + \mathcal{O}\left( \frac{2m_V}{\sqrt{s}} \right) \, ,
\end{align}
noting that this expansion is relevant also for the cross-section and the $R$-matrix~\cite{Romao:2016ien}.
Perturbative unitarity requires the bad high-energy behaviour to cancel and the $2\to 2$ amplitude to go at most as a constant in the high-energy limit ($s\sim -t \sim -u \gg M_V)$
\begin{align}
\lim_{\frac{\sqrt{s}}{2m_V} \rightarrow \infty} \hspace{-.75em} \mathcal{M} = \mathcal{M}^{(0)}, 
\end{align}
\ie, any amplitude respecting perturbative unitarity should have vanishing $\mathcal{M}^{(2)}$ and $\mathcal{M}^{(1)}$ at high energies.

\subsection{\texorpdfstring{$WW$}{WW} final state} 

Let us consider $e^+e^- \rightarrow W^+W^-$. We have four amplitudes contributing to the process, depicted in \cref{fig:diagrams_WW}:
$\mathcal{M}_\gamma$ and $\mathcal{M}_Z$ with an $s$-channel photon or $Z$-boson~(left), $\mathcal{M}_\nu$ with a $t$-channel neutrino~(centre), and $\mathcal{M}_h$ with a Higgs boson in the $s$-channel~(right).
The sum of the $\mathcal{M}_\gamma$, $\mathcal{M}_Z$, and $\mathcal{M}_\nu$ amplitudes grows with energy, displaying a unitarity violating behaviour.  
At high energies we find
\begin{align}
    \mathcal{M}_{\gamma+Z+\nu}(\pm\pm\rightarrow 00) \sim - \mathcal{M}_{h}(\pm\pm\rightarrow 00) \sim -\frac{e^2}{2\sin^2\theta_W} \frac{m_e}{m_W}\,\frac{\sqrt{s}}{2m_W} \, .
\end{align}
However, the Higgs exchange $\mathcal{M}_h$ has the same behaviour in the high-energy limit and exactly cancels the growth of the amplitude, recovering perturbative unitarity. As one might expect, this cancellation occurs only for the longitudinal final states, since the other cases do not grow with energy.
At the cross-section level, a similar cancellation occurs (see Ref.~\cite{Romao:2016ien} for an easy-to-access review). 

Let us now turn  our attention to the  Fano coefficients. As one might expect, the cancellation for the $\tilde{A}$ is the same as that happening at the cross-section level. At high energy and keeping the electron mass finite, the squared contribution of the sum of the $\gamma$, $Z$ and $\nu$ mediated diagrams to the Fano coefficient is
\begin{align}
\tilde{A}_{|\gamma+Z+\nu|^2} 
= \frac{e^4}{288\sin^4\theta_W}\frac{m_e^2}{m_W^2}\frac{s}{m_W^2} + \mathcal{O}(s^0)\,.
\end{align}
Adding the contribution of the Higgs diagram, we obtain three more terms,
\begin{align}
\tilde{A}_{|h|^2} \sim
-\tilde{A}_{(\gamma+Z+\nu)^*h} \sim
-\tilde{A}_{(\gamma+Z+\nu)h^*} \sim\tilde{A}_{|\gamma+Z+\nu|^2}
\quad\Longrightarrow\quad
\tilde{A}_{|\gamma+Z+\nu+h|^2}\sim\mathcal{O}(s^0)\,,
\end{align}
where now, due to the sign of the amplitudes, the interference terms of $(\gamma+Z+\nu)$ with the Higgs cancel with the matrix-element-squared contributions, resulting in a $\mathcal{O}(s^0)$ dependence.
This cancellation happens similarly for the $\tilde{a}_i$, $\tilde{b}_i$ and $\tilde{c}_{ij}$ coefficients. Specifically, we obtain the following non-zero results at order $s/m_W^2$ for the $(\gamma+Z+\nu)$ diagrams
\begin{align}
\tilde{a}_{3,|\gamma+Z+\nu|^2} \sim - \frac{1}{\sqrt{3}}\tilde{a}_{8,|\gamma+Z+\nu|^2} 
\sim \frac{e^4}{288\sin^4\theta_W}\frac{m_e^2}{m_W^2}\frac{s}{m_W^2}+\mathcal{O}(s^0) \,,
\end{align}
which in the case of the third component cancels when summed with the Higgs-mediated-squared and interference contributions, $\tilde{a}_{3,(\gamma+Z+\nu)^*h}+\tilde{a}_{3,(\gamma+Z+\nu)h^*}+\tilde{a}_{3,|h|^2}$, and likewise for the eighth component. 
The same expressions hold when considering $\tilde{b}_i$.
With respect to the correlation matrix $\tilde{c}_{ij}$, we have that for the $\gamma+Z+\nu$ diagrams
\begin{align}\begin{aligned}
\tilde{c}_{33,|\gamma+Z+\nu|^2} 
&= - \frac{1}{\sqrt{3}}\tilde{c}_{38,|\gamma+Z+\nu|^2}
= -\frac{1}{\sqrt{3}}\tilde{c}_{83,|\gamma+Z+\nu|^2}
= \frac{1}{3}\tilde{c}_{88,|\gamma+Z+\nu|^2} \\
&= \frac{e^4}{288\sin^4\theta_W}\frac{m_e^2}{m_W^2}\frac{s}{m_W^2}  +\mathcal{O}(s^0) \,.
\end{aligned}\end{align}
Adding the Higgs contribution leads to the same cancellation of all the coefficients at $s/m_W^2$ order, and perturbative unitarity is restored in the $R$-matrix. It is interesting to note that this cancellation occurs only for the coefficients of the third and eighth Gell-Mann matrix, while the others do not exhibit an energy growing behaviour. This occurs because in the high energy limit the $R$-matrix is dominated by the longitudinal polarisations and these are the Fano coefficients sensitive to those.

\subsection{\texorpdfstring{$ZZ$}{ZZ} final state}

We now turn to $ZZ$ production.
As in the previous case, let us focus on the lepton-initiated channel. The story for $ZZ$ is similar but now we only have two types of diagrams, $t$ and $u$-channel mediated by electrons and the Higgs $s$-channel. 
The cancellation in the amplitude occurs for the same helicities as before
\begin{align}
\mathcal{M}_{e}(\pm\pm \rightarrow 00) \sim \mathcal{M}_{h}(\pm\pm \rightarrow 00) \sim - e^2 \csc^2{2\theta}\,\frac{m_e}{2m_Z}\, \frac{\sqrt{s}}{2m_Z} + \mathcal{O}(s^0)\,,
\end{align}
where $\mathcal{M}_{e}$ represents both $t$ and $u$ channels. At the $R$-matrix level, the cancellation happens in a similar fashion to the $WW$ case. The $\tilde{A}$ coefficient for the electron diagrams,
\begin{align}
\tilde{A}_{e} = \frac{e^4}{288 \cos^4\theta_W\sin^4\theta_W}\frac{m_e^2}{m_Z^2}\frac{s}{m_Z^2}
+\mathcal{O}(s^0)\,,
\end{align}
cancel when adding the Higgs diagram. For the coefficients $\tilde{b}_i$ and $\tilde{c}_{ij}$, we again only have contributions to the third and eighth components and the cancellations repeat the previous pattern.

\subsection{Non-interference EFT effects}

The study of perturbative unitarity fits well in renormalisable theories where one does not expect growing amplitudes in the high-energy limit. This discussion is different for EFTs, in which higher-dimensional operators are included and they are allowed to grow with energy. This is actually a feature used in SMEFT analyses to probe deviations from the SM in tails of distributions. This growth can happen due to the particular new Lorentz structure of the operators or because of the spoiling of the SM cancellations. To understand the high-energy limit of the $R$-matrix and spin-related observables, let us first look at the amplitudes.

At high energy, the SM and the SMEFT induce a specific helicity pattern for diboson production. In \cref{tab:AmpsHighE}, as representative, we report the helicity amplitudes for $e^+e^- \to W^+W^-$. The helicity states are specified by the notation $\mathcal{M}(\lambda_1 \lambda_2| \alpha \beta)$, where $\lambda_1, \lambda_2$ are the helicities of the initial state electrons and $\alpha, \beta$ are the helicities of the final state EW bosons. We retain contributions up to order $\mathcal{O}(x^0)$  (see also Refs.\cite{Falkowski:2016cxu} and~\cite{Helset:2017mlf}).

\begin{table}[!ht]
\centering
\renewcommand{\arraystretch}{1.4}
\small
\begin{tabular}{cccc}
\toprule
 $(\lambda_1 \lambda_2| \alpha\, \beta)$ & \text{SM}   & EFT $\Lambda^{-2}: c_{WWW}$  \\ \midrule
 $+-00$ & $-2\sqrt{2}G_F m_Z^2\sin\theta$ &   - \\ 
 $+--+$  & $2\sqrt{2}G_F m_W^2\sin\theta$  &   -  \\
 $+-+-$  & $-\frac{1}{\sqrt{2}}G_F m_W^2\sin^3\theta\csc^4(\theta/2)$  &  -   \\
 $+-\pm\pm$ &- & { $3\cdot 2^{1/4} \sqrt{G_F} m_W\sin\theta \, ( 4 m_W^2 x^2 - m_Z^2)$ }\\
 $+-0\pm$ &- & {$-3\cdot 2^{3/4} \sqrt{G_F}m_W^3(\pm 1+ \cos\theta) \, x$}\\
$+-\pm 0$ &- & {$-3\cdot 2^{3/4} \sqrt{G_F}m_W^3(\mp 1+ \cos\theta) \, x$}\\
  \hline
 $-+00$  & $2\sqrt{2}G_F (m_Z^2-m_W^2)\sin\theta$  & -  \\
$-+\pm\pm$ & - & $6\cdot2^{1/4}\sqrt{G_F}m_W(m_Z^2-m_W^2)\sin\theta$  \\ 
\hline\bottomrule
\end{tabular}
\caption{Helicity pattern in the high-energy limit of electron-initiated diboson production amplitudes for the SM and the SMEFT amplitudes induced by the $\mathcal{O}_W$ operator, which has a distinguished Lorentz structure. The $\pm\pm\alpha\beta$ case has all zero entries, regardless of $\alpha\beta$. The short-hand notation $x=\sqrt{s}/(2m_W)$ is used. Contributions to the helicity amplitude that are sub-leading and energy suppressed are indicated with -.}
\label{tab:AmpsHighE}
\end{table}

It is clear that the SM and the $\mathcal{O}_W$ operator induce different helicity amplitudes. When computing a cross-section, where the on-shell final states are the gauge bosons, this leads to a vanishing EFT linear correction due to non-interference of the amplitudes. The cancellation for massless particles can be proven by applying helicity selection rules, see Ref.~\cite{Azatov:2016sqh}. However, one can show that the interference can be recovered exploiting the angular distributions of the decay products, e.g.\ considering the full process $e^+e^- \rightarrow VV \rightarrow 4f$, with $f$ either a lepton or a quark. For in-depth phenomenological studies of this aspect we refer the reader to the literature~\cite{Panico:2017frx,Franceschini:2017xkh,Azatov:2017kzw,Helset:2017mlf,Azatov:2019xxn,Aoude:2019cmc,Degrande:2021zpv}.

In the $R$-matrix formulation, this translates into the fact that the diagonal terms at the linear EFT level vanish, while the off-diagonal terms allow for a resurrection of the interference between the SM and the operator $\mathcal{O}_W$. In the high energy and massless limit, the Fano coefficient $\tilde{A}$ has a vanishing linear EFT contribution
\begin{align}
\tilde{A}(\mathcal{O}_{W}) \sim 0 \, \, ,
\end{align}
but the other Fano coefficients can be different from zero and potentially allow for increased sensitivity. Defining $x=\sqrt{s}/(2m_V)$ and in the limit $m_W\sim m_Z \sim m_V$ we have
\begin{subequations}
    \begin{align}
    \tilde{a}_1 (\mathcal{O}_{W}) &\simeq\tilde{b}_1(\mathcal{O}_{W}) \simeq \bar{c}_{W} \, 2^{5/4} \, x \; \cos^4(\theta/2)(\cos \theta+3)\csc \theta \, ,\\
    \tilde{a}_4 (\mathcal{O}_{W}) &\simeq\tilde{b}_4(\mathcal{O}_{W}) \simeq - \bar{c}_{W} \, 2^{3/4} (\cos\theta ((4x^2-3)\cos \theta+4x^2+1)+2)\, ,\\
    \tilde{a}_6 (\mathcal{O}_{W}) &\simeq\tilde{b}_6(\mathcal{O}_{W}) \simeq \bar{c}_{W} \, 2^{1/4} \, x \,\sin^2(\theta/2)\sin \theta \, ,
\end{align}
\end{subequations}
with $\bar{c}_{W}=c_{W} \, G_F^{3/2} \, m_V^5 $. The spin-spin Fano coefficients $\tilde c_{ij} (=\tilde c_{ji})$ for the operator $\mathcal{O}_{W}$ are given by 
\begin{subequations}
\begin{align}
\tilde{c}_{13}
&\simeq 
3 \, \bar{c}_{W} \cdot 2^{3/4}  \cos^2(\theta/2)(3\cos\theta+1)\cot(\theta/2)\, x\\
\tilde{c}_{14}
&\simeq 
-\tilde{c}_{25}
\simeq 
\tilde{c}_{46}
\simeq
-\tilde{c}_{57}
\simeq 
-3 \, \bar{c}_{W} \cdot 2^{3/4} \sin\theta(1+\cos\theta)\, x \, ,\\
\tilde{c}_{16} &\simeq \tilde{c}_{27} \simeq 3 \, \bar{c}_{W} \cdot 2^{-5/4} \sin^2(\theta/2)((3-4x^2)\cos\theta-4x^2+1) \, ,\\
\tilde{c}_{18} &\simeq 
\bar{c}_{W} \, \sqrt{3} \, \cdot 2^{-3/4} \cos^2(\theta/2) (\cos\theta+2)\cot(\theta/2)\, x \, ,\\
\tilde{c}_{35} &\simeq 
\bar{c}_{W} \, \sqrt{3} \, \cdot 2^{-3/4} \sin^2(\theta/2) \sin\theta \,  x \, ,
\\
\tilde{c}_{48} &\simeq 
\sqrt{3} \, \bar{c}_{W} \cdot   2^{-9/4} (8(1-x^2)\cos\theta+(4x^2-3)\cos(2\theta)- 5(4x^2-1)) \, ,\\
\tilde{c}_{68} &\simeq 
- 5\sqrt{3} \, \bar{c}_{W} \cdot 2^{-3/4}\sin^2(\theta/2)\sin\theta  \, .
\end{align}
\end{subequations}
The omitted ones are vanishing, showing that this matrix is rather sparse and several of the energy growing terms are closely related between each other. Here, we note also that the relevant coefficients for the perturbative unitarity cancellation vanish in the non-interference analysis, as they are pertinent to the longitudinal polarisations, while the $\mathcal{O}_W$ induces energy growing behaviour in transverse helicity configurations.

\section{Entanglement at particle colliders}
\label{sec:Entanglement_colliders}

In the following we will study the spin density matrix and the presence of entanglement in diboson production for both lepton and proton colliders. In the case of the lepton collider, we consider centre of mass energies up to \SI{1}{\TeV}, while in the case of the proton collider we focus on the LHC setup with a centre of mass energy of \SI{13}{\TeV}. In the latter we work in the $5$ flavour scheme, \ie, all quarks are massless aside for the top quark, and we use the latest \texttt{NNPDF4.0} NNLO PDF set~\cite{Ball_2022}. The input parameters used for the calculations are the following
\begin{align*}
    m_W &= \SI{80.377}{\GeV} \, ,  & m_Z &= \SI{91.1876}{\GeV} \, , \\
    m_h &= \SI{125.35}{\GeV} \, ,  & G_F &= \SI{1.1663788e-5}{\GeV^{-2}} \, .
\end{align*}

Calculations are performed analytically, taking advantage of the \texttt{Feynrules}~\cite{Darme:2023jdn,Alloul_2014}, \texttt{FeynArts}~\cite{Hahn_2001} and \texttt{FeynCalc}~\cite{Shtabovenko:2020gxv} tool chain. Amplitude computations are also validated numerically with \texttt{MadGraph5\_aMC@NLO}~\cite{Alwall:2014hca} and \texttt{SMEFT@NLO}~\cite{Degrande:2020evl}. The results are presented in the centre of mass frame of the diboson pair, as function of the kinematical variables $m_{VV}$, the invariant mass of the system, and $\cos{\theta}$, the angle between the initial state anti-particle and the $W^+$ or $Z$ boson. In the case of the SM calculations, we have validated our results against the ones obtained in Ref.~\cite{Fabbrichesi:2023cev}, finding excellent agreement.

For each process considered, we computed the density matrix including EFT corrections (both linear and quadratic) and consequently the various Fano coefficients, from which we determine analytically the entanglement related markers \cMB and \cUB, the purity~$P$ and the indicator $\langle \mathcal{B}\rangle_\mathrm{max}$ for Bell inequality violation.\footnote{%
    The minimisation in the calculation of the latter is, however, done numerically.
}

\subsection{Lepton collider}

In this section we discuss the entanglement pattern in diboson production at a lepton collider, focusing in particular on its dependence on the couplings in the context of the SMEFT. In a lepton collider, the collision energy is fixed. This means that at LO, without considering initial state QED (or more in general EW) radiation, the diboson pair is produced with an invariant mass $m_{VV}$ which is identical to the initial state energy. In this setup, we can therefore study the behaviour of the entanglement by fixing the collider energy and looking at its angular dependence. This remains true if we consider lepton colliders with energies up to a few \si{\TeV}. In the scenario of a multi-\si{\TeV} lepton collider, such as a circular muon collider, the machine would effectively behave as an EW boson collider and the main mode of production for diboson pair will be vector boson scattering~\cite{Costantini:2020stv}, see also Refs.~\cite{Al_Ali_2022,Accettura:2023ked,Aime:2022flm,MuonCollider:2022xlm,Chiesa:2020awd} for in depth studies of muon collider physics. The study of this kind of processes, which would be phenomenologically quite different and would add considerable sensitivity on dimension-8 operators, is left for future work.

\subsubsection*{$WW$ production}

\begin{figure}[t!]
    \centering
    \includegraphics[width=.9\linewidth]{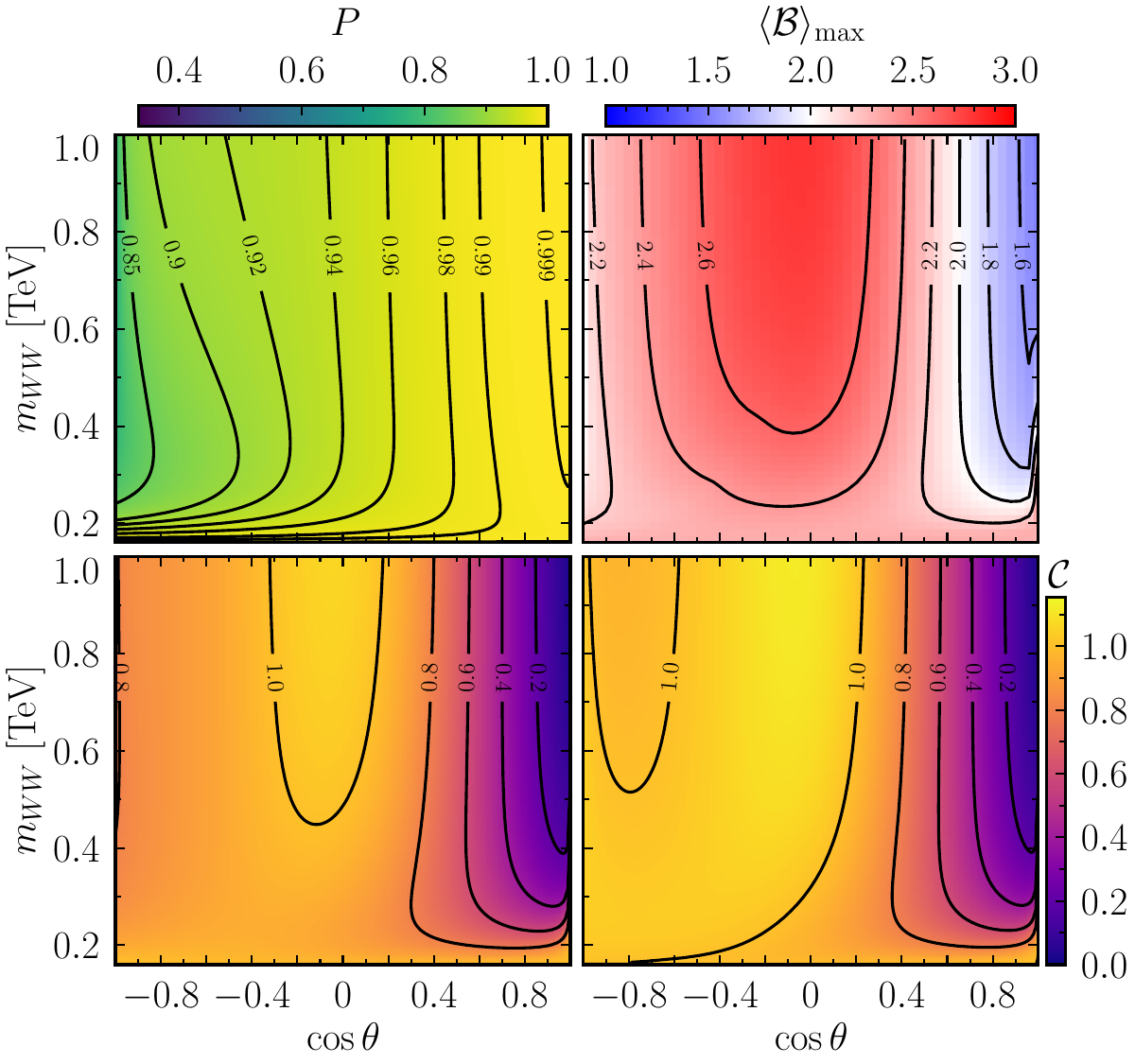}
    \caption{%
        Entanglement in the $e^+ e^- \to W^+ W^-$ channel in the SM. We show the lower bound~\cMB~(bottom left) and upper bound~\cUB~(bottom right) on the concurrence \concurrence, the purity~$P$~(top left) as well as the indicator~$\langle\mathcal{B}\rangle_\mathrm{max}$ for Bell inequality violation~(top right) as a function of the invariant mass $m_{WW}$ (or equivalently the collider energy) and the cosine of the angle between the positron and the $W^+$ in the centre of mass frame.
    }
    \label{fig:entang_eeWW}
\end{figure}
\begin{figure}[t!]
    \centering
    \includegraphics[width=.45\linewidth]{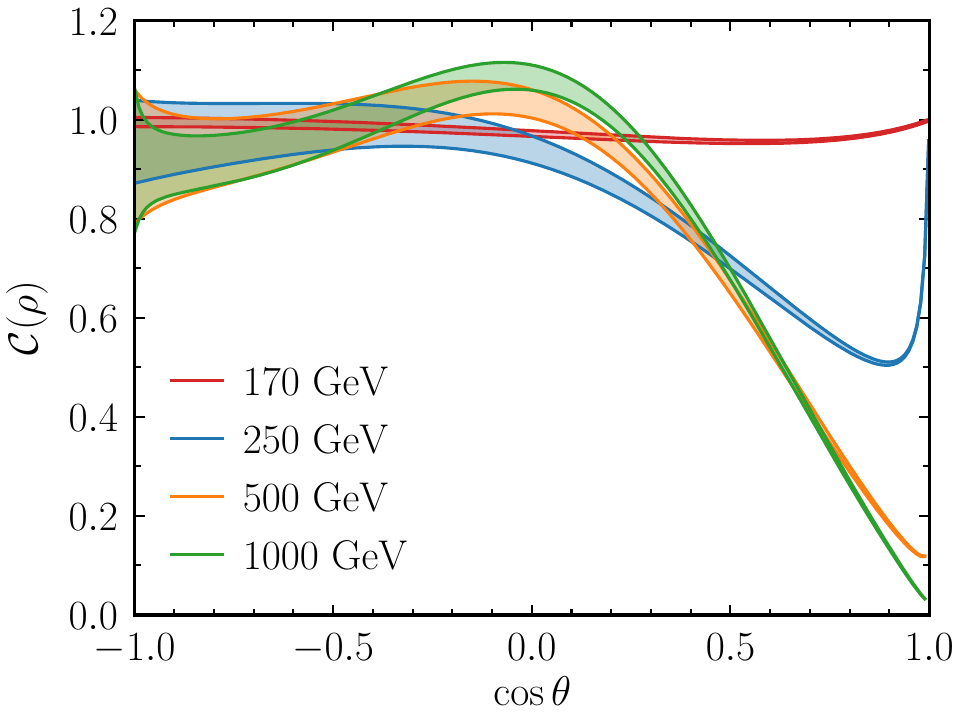}
    \includegraphics[width=.45\linewidth]{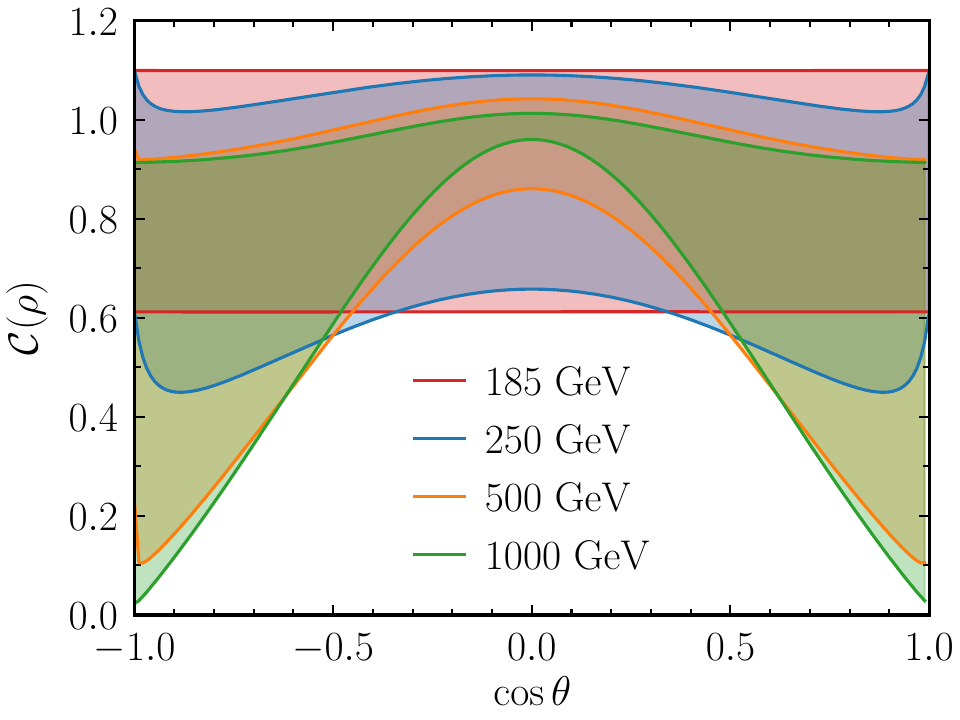}
    \caption{For benchmark fixed collider energies, we show the expected value for the concurrence as a function of $\cos{\theta}$ in the SM. The bands are determined by the lower and upper bounds of the concurrence, \ie, \cMB and \cUB. Left: $e^+ e^- \to W^+ W^- $. Right: $e^+ e^- \to Z Z$. }
    \label{fig:bands_ee}
\end{figure}

The production of a pair of $W$ bosons is of most interest in a lepton collider, as it is characterised by a considerably high cross section even at high energies. For instance, at \SI{1}{\TeV}, the total cross section is still of the order of a few \si{\pico\barn}. This means that the process is going to be particularly advantageous in terms of statistics and could prospectively allow to detect entanglement with strong significance~\cite{Fabbrichesi:2023cev}.

The relevant couplings for the process are the coupling of the lepton to the $W$ boson $g_W$, the coupling to the $Z$ boson $g_V^Z$ and $g_A^Z$, and the TGCs. The process is therefore sensitive to several of the EW interactions.
Furthermore, in the SMEFT framework strong correlations are present among the parameters.

In \cref{fig:entang_eeWW} we show the entanglement pattern we can expect from the production of a pair of $W$ bosons in the SM. Results are shown as a function of the kinematical variables $m_{WW}$ and $\cos{\theta}$. Note that there is no symmetry around $\theta = \pi/2$, as the EW interactions are not parity invariant. This is not the case when considering for instance $t\bar{t}$ production, where the process is dominated by QCD~\cite{Aoude:2022imd}.
In the upper left figure we show the values for the purity indicator~$P$, which depicts a scenario where the majority of the phase space is characterised by a density matrix close to maximal purity.
In the bottom left and bottom right figure, we show the value of the lower and upper bound for the concurrence.
The plots demonstrate that the diboson pair has in general a very high value of the concurrence across the phase space, indicating that entanglement is present almost everywhere, with the exception of the collinear region with $\theta = 0$, where low values of entanglement are expected. 

This can be seen more explicitly in the left plot of \cref{fig:bands_ee}, where four collider energies, 170, 250, 500 and 1000 GeV, are chosen as benchmarks. The plot displays bands for the concurrence, where the lower and the upper bound are given by \cMB and \cUB respectively, as a function of $\cos{\theta}$. Entanglement is high and stable for most of the angles, but decreases sharply as we approach the forward collinear limit at high energies. Also, the closer we are to threshold the tighter the bands get, indicating that the quantum state of the process goes towards maximal purity, as confirmed from the purity plot in \cref{fig:entang_eeWW}. In fact, as previously discussed, in this limit \cMB and \cUB coincide.

Finally, in the upper right plot of \cref{fig:entang_eeWW} we display the Bell inequality violation indicator~$\langle\mathcal{B}\rangle_\mathrm{max}$. The pattern in the figure closely resembles that shown for the concurrence marker~\cMB and as expected, the violation is ubiquitous. In particular, Bell inequalities are severely violated when high values of entanglement are present, \ie,  at high energy in the central region.\footnote{Note that we validated our results against Ref.~\cite{Fabbrichesi:2023cev} finding excellent agreement for most of phase space but some discrepancy in the forward region $\cos{\theta} = 1$, where the authors find a slight violation of Bell inequalities, \ie $\langle\mathcal{B}\rangle_\mathrm{max} > 2$. However, as explicitly discussed above, we find that in the phase space region in question the density matrix of the system is described by a pure separable state. Therefore, no Bell inequality violation is expected, in agreement with \cref{fig:entang_eeWW}. This property has been also verified with a numerical simulation in \texttt{MadGraph5\_aMC@NLO}, finding good agreement with the analytical calculation.}

In order to gain a better insight, it is useful to see how the density matrix decomposes in terms of quantum states at particular phase space points. For instance, we find that the diboson pair is produced at threshold in a pure and entangled quantum state, \ie, 
\begin{equation}
    \ket{\Psi(m_{WW} = 2 \, m_W)} = \frac{1}{\sqrt{2}} \left( \ket{+ 0}_{\boldsymbol{p}} + \ket{0 +}_{\boldsymbol{p}} \right) = \ket{\Psi_{0+}}_{\boldsymbol{p}} \, ,
\end{equation}
where $\ket{s_1 s_2}_{\boldsymbol{p}} = \ket{s_1}_{\boldsymbol{p}} \otimes \ket{s_2}_{\boldsymbol{p}}$ and $\ket{s}_{\boldsymbol{p}}$ is the eigenstate of the spin operator in the direction of the beam line~$\boldsymbol{p}$ with eigenvalue $s$. This means  that at threshold, the diboson pair is produced in an entangled state, characterised by total spin $2$ and spin component along $\boldsymbol{p}$ equal to $1$. However, despite the fact that this is an entangled state, the value of the concurrence at threshold is $1$, not reaching the maximal value of $2/\sqrt{3}$. We find that this quantum state is unaffected by the presence of new physics. Even when EFT effects are taken into account, the quantum state of the system is still expressed by $ \ket{\Psi_{0+}}_{\boldsymbol{p}}$. The reason for that is that most of the contributions are identically zero at threshold and the ones that are not, are simply shifting the absolute value of $g_W$, resulting in an increased total cross section but not affecting the spin correlation patterns. It turns out indeed that at threshold, the only relevant coupling for the process is $g_W$.
Equally interesting is the situation at high energy.
In particular, in the central region, we find that the density matrix is characterised by a mixed quantum state but dominated by the presence of a pure entangled quantum state, which explains the high concurrence. Specifically, the density matrix can be defined with respect to the $\boldsymbol{k}$ direction, the momentum of the $W^+$ boson, in the following way
\begin{equation}
    \rho(m_{WW} \to \infty, \cos{\theta} = 0) = p_1 \, \ket{1}\bra{1} + p_2 \, \ket{2}\bra{2}\, ,
\end{equation}
with
\begin{align}\begin{aligned}
    \ket{1} &= 0.64 \ket{++}_{\boldsymbol{k}} - 0.64 \ket{--}_{\boldsymbol{k}} + 0.43 \ket{00}_{\boldsymbol{k}} \, ,\\
    \ket{2} &= 0.3 \ket{++}_{\boldsymbol{k}} - 0.3 \ket{--}_{\boldsymbol{k}} - 0.9 \ket{00}_{\boldsymbol{k}} \, ,
\end{aligned}\end{align}
and $p_1 \approx 0.97$ and $p_2 \approx 0.03$.

On the other hand, in the collinear region, $\theta = 0$, the concurrence goes to zero (see \cref{fig:entang_eeWW}) and we find that the diboson pair is produced in a separable state, \ie
\begin{equation}
    \ket{\Psi(m_{WW} \to \infty, \cos{\theta} = 1)} = \ket{+ +}_{\boldsymbol{k}} \, .
\end{equation}

\begin{figure}[t!]
    \centering
    \includegraphics[width=.9\linewidth]{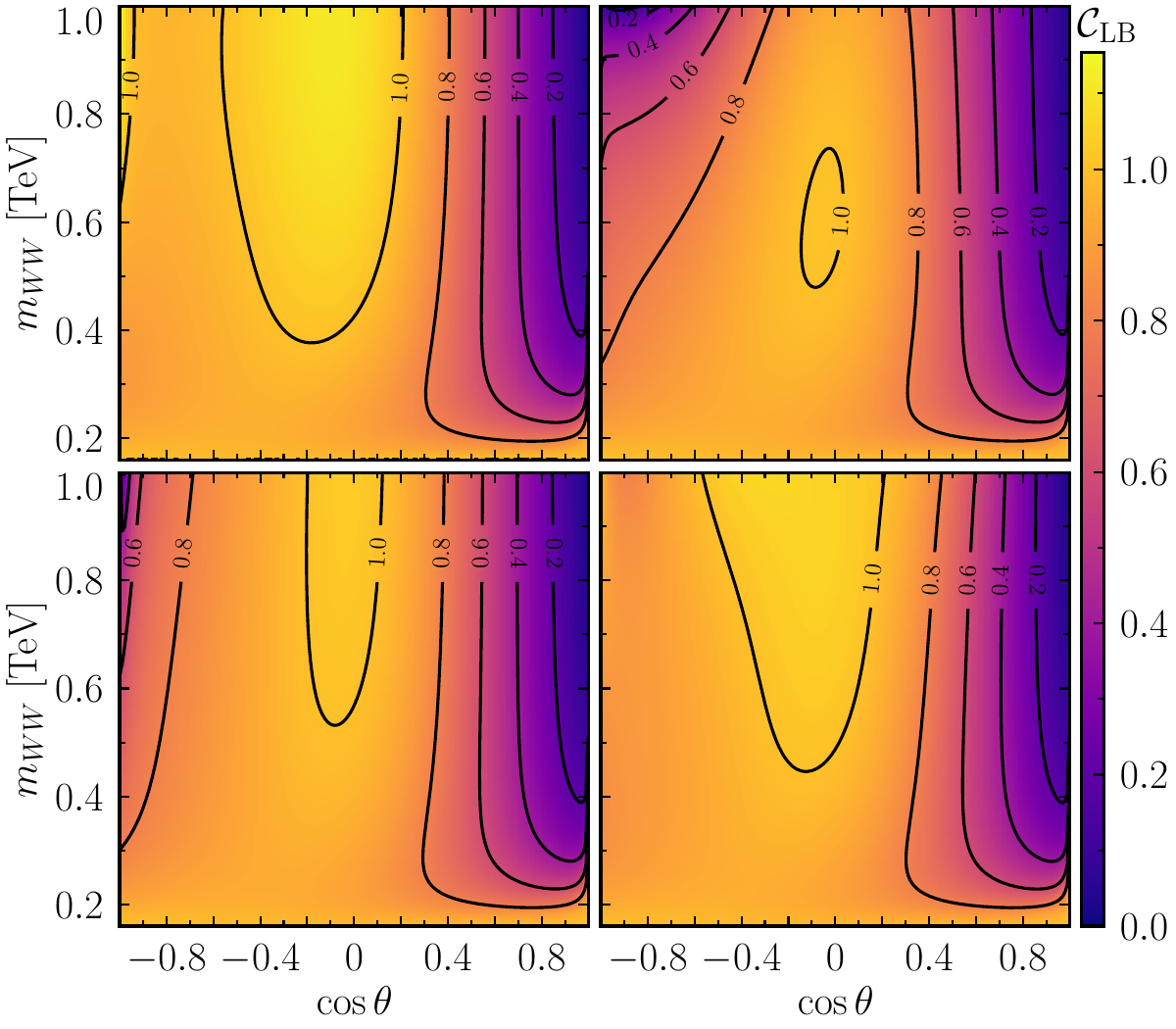}
    \caption{The changes in the marker \cMB is shown for a selection of operators and benchmark Wilson coefficient values for the production of $W^+W^-$ at a lepton collider. Only one operator at the time is switched on. Top left: $c_{\varphi e} = \SI{0.1}{\TeV^{-2}}$, top right: $c_{\varphi l}^{(1)} = \SI{0.1}{\TeV^{-2}}$, bottom left: $c_{\varphi WB} = \SI{0.25}{\TeV^{-2}}$, bottom right: $c_{W} =\SI{0.25}{\TeV^{-2}}$.}
    \label{fig:entang_eeWW_EFT}
\end{figure}

We now move to discuss the effects of dimension-6 operators on the spin density matrix. In \cref{fig:entang_eeWW_EFT}, we show the effects of a selected number of operators on the marker \cMB. For each operator we choose values that are currently allowed or at the boundary of the most up to date global fit studies (see \cref{tab:ops} and for example Refs.~\cite{Ethier:2021bye, Ellis:2020unq}). In the context of the EFT, here and in the following sections, we decide to limit ourselves mostly to show the effects only for the \cMB indicator, as a good representative metric for the entanglement pattern across phase space. In some cases, we also show the purity for comparison. Note however that the calculation performed and the expressions provided in the ancillary files allow for a complete determination of the density matrix in the SMEFT and consequently for the calculation of every quantum observable derived from it.
We show the effects of the operators by turning on only one of them at the time, in order to display how different modifications of the couplings alter the entanglement pattern. It is particularly interesting to see that not only the effects of the operators are substantial for the value of choice of the Wilson coefficients, but that the pattern of modification severely changes from operator to operator. The ultimate reason for that has to be tracked down to the way they induce shifts to the EW couplings defined in \cref{sec:interactions}. For instance, the operator $\OO_{ll}$ (not displayed) is affecting the SM couplings in a universal way, inducing an overall rescaling factor, and therefore does not affect the density matrix in any way, leaving the \cMB marker unchanged. On the other hand, the four Wilson coefficients shown in \cref{fig:entang_eeWW_EFT}, $c_{\varphi e}$, $c_{\varphi l}^{(1)}$, $c_{\varphi WB}$ and $c_{W}$, alter the EW interactions in such a way that the density matrix of the final state spins is sensibly affected. We can see for instance that a positive value of $c_{\varphi e}$, which shifts the value of the right-handed coupling to the $Z$ boson, induces an augmented level of entanglement at high energy both in the central region and in the backward one. Very different behaviour is instead produced by a positive $c_{\varphi l}^{(1)}$, which by shifting the left-handed coupling to the $Z$ boson decreases the value of the \cMB marker in the same region.
The effects of the $c_{\varphi WB}$ and $c_{W}$ operators is instead milder. In particular, one could have potentially expected a big impact from the $\OO_{W}$ operator given that it induces the presence of the non-SM Lorentz structure $\delta \lambda_V$, but that does not seem to be the case. Note that in general, to an opposite sign value of the Wilson coefficient corresponds an opposite effect, \ie,  if for $c_{W}=\SI{0.25}{\TeV^{-2}}$ the high entanglement region increases, for $c_{W}= \SI{-0.25}{\TeV^{-2}}$ we would see a decrease. 
Finally, it is worth noticing that all of the operators leave the entanglement pattern in the forward region unchanged and have mostly sensible effects in the high energy region, as one would have naively expected. 

\begin{figure}[t!]
    \includegraphics[width=\linewidth]{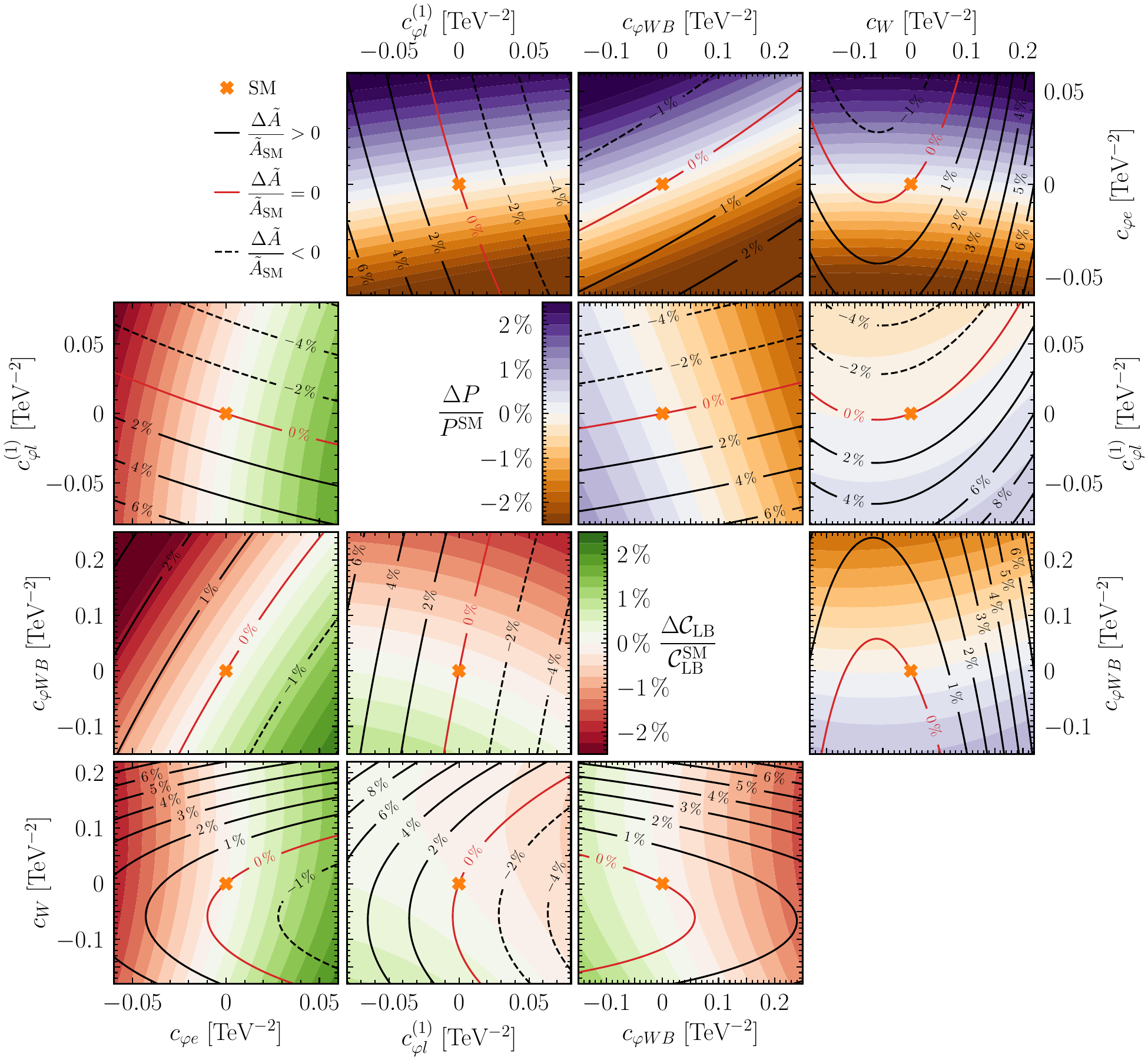}
    \caption{%
        The relative changes in the marker \cMB~(lower triangle) and purity~$P$~(upper triangle) compared to the SM values $\cMB^\mathrm{SM}=1.0$ and  $P^\mathrm{SM}=0.94$ as a function of the Wilson coefficients $c_{\varphi e}$, $c_{\varphi l}^{(1)}$, $c_{\varphi WB}$ and $c_{W}$ for $W^+W^-$ production at a lepton collider, at $m_{WW}=\SI{500}{\GeV}$ and $\cos\theta=0$.
        The lines indicate the relative change of the cross section.
    }
    \label{fig:entang_eeWW_EFT_triangle}
\end{figure}

Finally, in \cref{fig:entang_eeWW_EFT_triangle}, we depict the relative change of \cMB~(lower triangle) and the purity~(upper triangle) as a function of the Wilson coefficients, varying two coefficients at a time and considering the fixed phase space point $m_{WW}=\SI{500}{\GeV}$ and $\theta=\pi/2$. 
Here, $\Delta\cMB$ and $\Delta P$ denote the difference between the marker~\cMB and the purity, respectively, calculated within the SMEFT, and the SM values, $\cMB^\mathrm{SM}=1.0$ and $P^\mathrm{SM}=0.94$. The SMEFT values are calculated including dimension-6 and dimension-6 squared contributions.
In addition, the contours depict the relative change in $\tilde{A}$, \ie,  the relative change of the differential cross-section with respect to the SM.
Notably, we see that that the spin-related observables generally probe different parameter directions than the cross-section, potentially offering complementary probes of NP. This could be of fundamental importance both for discovery, enhancing the sensitivity to EFT corrections, and for characterisation in the event of a clear deviation from the SM. Additionally, one can clearly see from the $c_{W}$ plots that the spin-related observables display a resurrection of the interference, as expected, while the differential cross-section contour lines are mostly dominated by quadratic corrections.

\subsubsection*{$ZZ$ production}

\begin{figure}[t!]
    \centering
    \includegraphics[width=.9\linewidth]{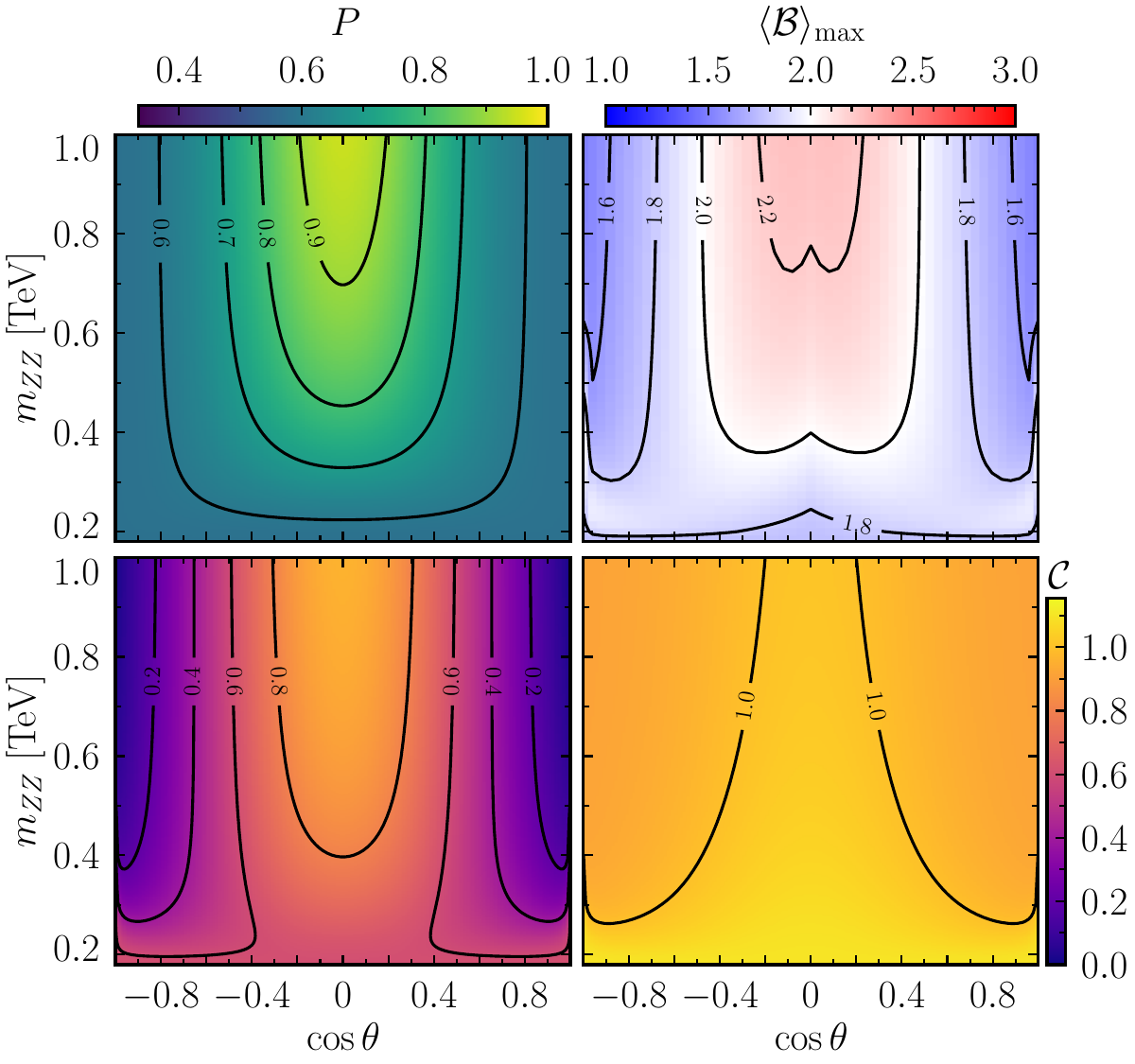}
    \caption{%
        Entanglement in the $e^+ e^- \to Z Z$ channel in the SM. We show the lower bound~\cMB~(bottom left) and upper bound~\cUB~(bottom right) on the concurrence \concurrence, the purity~$P$~(top left) as well as the indicator~$\langle\mathcal{B}\rangle_\mathrm{max}$ for Bell inequality violation~(top right) as a function of the invariant mass $m_{ZZ}$ (or equivalently the collider energy) and the cosine of the angle between the positron and the $Z$, in the centre of mass frame.
    }
    \label{fig:entang_eeZZ}
\end{figure}

One key difference between $Z Z$ and $W^+ W^-$ production is the fact that they probe complementary couplings of the fermions to the diboson system. In particular, in the case of $Z Z$, only the coupling of the fermions to the $Z$ boson are relevant, while in the case of $W^+ W^-$ a more intricate coupling dependence is present, including the triple gauge coupling. For $Z$ pair production, the scattering amplitudes are completely determined by the values of the vectorial and axial couplings to the $Z$ boson in \cref{eq:Z_coupl}.

In \cref{fig:entang_eeZZ} we show the entanglement pattern in the SM. Contrary to the $W^+ W^-$ case, we see a symmetry with respect to $\cos{\theta}=0$, given that this time the final state has identical particles and consequently the system exhibits a symmetry under parity transformations. 
The plot in the upper left corner further indicates that, in contrast to $W^+ W^-$ production, the majority of the phase space for $ZZ$ production is characterised by a mixed quantum state. High purity $P$ is reached only in the high energy central region.
The plot on the lower left depicts the entanglement pattern in terms of the lower bound marker \cMB. According to that, the entanglement is expected to be high at high energy in the central region, but quite low in the forward region. However, the picture gets more complicated if we look at the right panel of \cref{fig:bands_ee}, which displays bands for the concurrence,
making use of both the lower and the upper bound, for benchmark collider energies as a function of $\cos{\theta}$. In the figure, we see that the lower bound goes towards zero in the collinear regime, but the upper bound does not, giving us enormous uncertainty on the determination of the entanglement with this approach. This is also confirmed by the plot in the lower right panel of \cref{fig:entang_eeZZ} displaying the upper bound marker \cUB across phase space. Almost all of the phase space is characterised by \cUB close to maximal.
Finally, in the upper right plot of \cref{fig:entang_eeZZ} we report on the expected value for the Bell inequality violation marker $\langle\mathcal{B}\rangle_\mathrm{max}$. In contrast to the $W^+ W^-$ final state, the region of phase space with $\langle\mathcal{B}\rangle_\mathrm{max} > 2$ is rather limited and slight violations are only present in the high energy central region. 

More information on the entanglement can be gathered by directly inspecting the density matrix and its decomposition in terms of quantum states. In particular, we find that at threshold the $Z$ boson pair is produced in a mixed state
\begin{equation}
    \rho(m_{ZZ} = 2 \, m_Z) = p_1 \, \ket{\Psi_{0+}}_{\boldsymbol{p}}\bra{\Psi_{0+}}_{\boldsymbol{p}} + p_2 \, \ket{\Psi_{0-}}_{\boldsymbol{p}}\bra{\Psi_{0-}}_{\boldsymbol{p}}\, ,
\end{equation}
with $p_1 = 0.7$ and $p_2 = 0.3$ and 
\begin{align}\begin{aligned}
    \ket{\Psi_{0+}}_{\boldsymbol{p}} &= \frac{1}{\sqrt{2}} \left( \ket{+ 0}_{\boldsymbol{p}} + \ket{0 +}_{\boldsymbol{p}} \right) \, ,\\
    \ket{\Psi_{0-}}_{\boldsymbol{p}} &= \frac{1}{\sqrt{2}} \left( \ket{- 0}_{\boldsymbol{p}} + \ket{0 -}_{\boldsymbol{p}} \right) \, .
\end{aligned}\end{align}
The two states are both fully entangled, but, given the fact that the density matrix is in a mixed state, the value of the concurrence is not maximal.

\begin{figure}[t!]
    \centering
    \includegraphics[width=.9\linewidth]{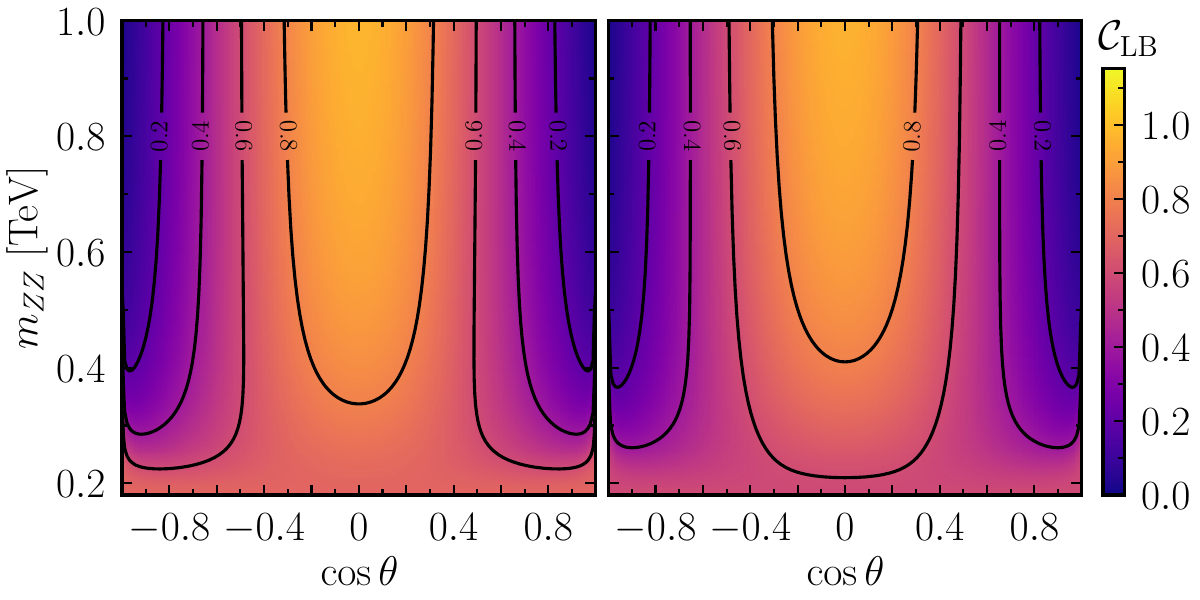}
    \caption{The changes in the marker \cMB is shown for a selection of operators and benchmark Wilson coefficient values for the production of $ZZ$ at a lepton collider. Only one operator at the time is switched on. Left: $c_{\varphi D} = \SI{0.5}{\,TeV^{-2}}$, right: $c_{\varphi l}^{(3)} = \SI{-0.25}{TeV^{-2}}$.}
    \label{fig:entang_eeZZ_EFT}
\end{figure}

\begin{figure}[t!]
    \centering
    \includegraphics[width=.9\linewidth]{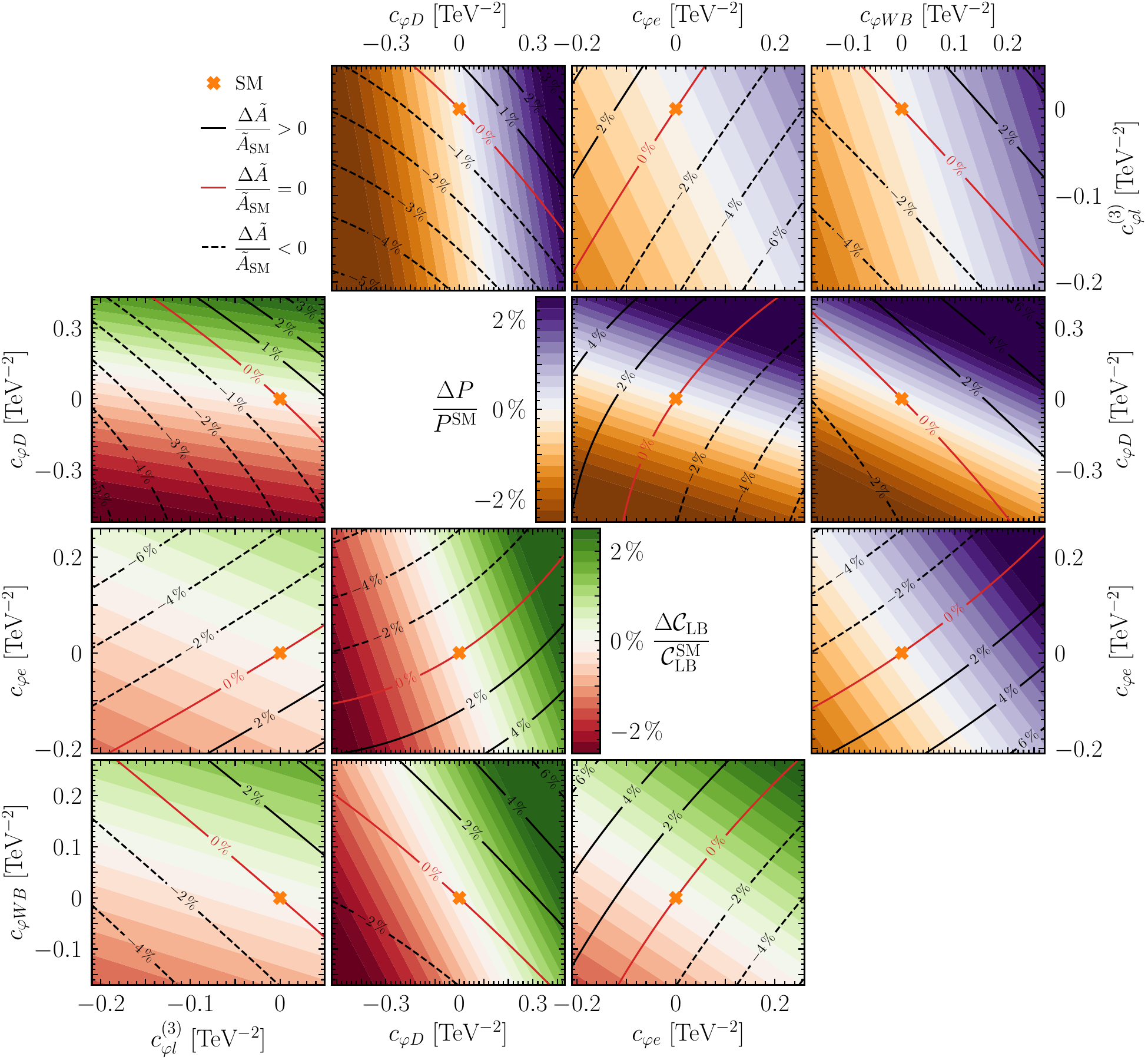}
    \caption{%
        The relative changes in the marker \cMB~(lower triangle) and purity~$P$~(upper triangle) compared to the SM values $\cMB^\mathrm{SM} = 0.86$ and $P^\mathrm{SM}=0.83$ as a function of the Wilson coefficients $c_{\varphi l}^{(3)}$, $c_{\varphi D}$, $c_{\varphi e}$ and $c_{\varphi W B}$ for $Z Z$ production at a lepton collider, at $m_{ZZ}=500\,\text{GeV}$ and $\cos\theta=0$.
        The lines indicate the relative change of the cross section.
    }
    \label{fig:entang_eeZZ_EFT_triangle}
\end{figure}

On the other hand, at high energy the picture is different. In the central region, $\theta = \pi/2$, 
the diboson pair is produced in a pure spin-2 maximally entangled state
\begin{equation}
    \ket{\Psi_{+-}}_{\boldsymbol{k}} = \frac{1}{\sqrt{2}} \left( \ket{+ +}_{\boldsymbol{k}} - \ket{- -}_{\boldsymbol{k}} \right) \, .
\end{equation}
This is fully consistent with what we observe in \cref{fig:entang_eeZZ}.
The concurrence study based on the lower and upper bounds is inconclusive in the forward region. On the other hand, by inspecting the density matrix directly at $\theta=0$ in the high energy limit, we find that the $Z$ pair is produced in a mixed ensemble of separable quantum states, \ie
\begin{equation}
    \rho(m_{ZZ} \to \infty, \cos{\theta} = 1) = p_1 \, \ket{+ +}_{\boldsymbol{p}}\bra{+ +}_{\boldsymbol{p}} + p_2 \, \ket{- -}_{\boldsymbol{p}}\bra{- -}_{\boldsymbol{p}}\, ,
\end{equation}
with $p_1=0.7$ and $p_2=0.3$. We are therefore able to conclude that in the forward region, the diboson pair is indeed not entangled as suggested by the \cMB marker behaviour.

As for the case of $W^+ W^-$ production, in \cref{fig:entang_eeZZ_EFT} we show the effects of a selected number of operators on the marker \cMB. Since for $ZZ$  production the dependence on the couplings is considerably simpler, the only defining parameter of the entanglement pattern is the balance between the vector and the axial coupling, or, to be more precise, the ratio between the two. We observe that, aside for the operators that universally rescale both couplings leaving the density matrix unchanged, the behaviour of all the operators is phenomenologically identical. In the figure, we show the two possible deviating patterns from the SM, choosing as benchmark Wilson coefficients $c_{\varphi D} = \SI{0.5}{\TeV^{-2}}$ and $c_{\varphi l}^{(3)} = \SI{-0.25}{TeV^{-2}}$. In the case of the former, the entanglement marker is augmented almost everywhere, indicating that the Wilson coefficient shifts the couplings in such a way that the dominant coupling prevails a bit more and the density matrix is slightly ``less mixed''. On the other hand, for $c_{\varphi l}^{(3)} = \SI{-0.25}{\TeV^{-2}}$ we observe a decrease of the entanglement across the board. Note that at variance with  $W^+ W^-$ production, the effects close to threshold are a bit more pronounced. We stress again that none of these behaviours is pertinent to a specific operator, but both the increase and decrease of entanglement can be produced by any of the operators by switching on the corresponding Wilson coefficient (with negative and positive values for opposite effects).
Finally, in \cref{fig:entang_eeZZ_EFT_triangle} we show the relative change of \cMB and the purity as a function of a selection of Wilson coefficients,  considering the fixed phase space point $m_{ZZ}=\SI{500}{GeV}$ and $\theta=\pi/2$. As for the case of $W$ pair production, the complementarity between differential cross section and spin-related observables is self-evident. Additionally, it is interesting to see that purity and \cMB probe the same direction in the Wilson coefficient parameter space, indicating that there is a strong correlation between the two observables at the spin density matrix level.

\subsection{Hadron collider}

We now present results for a hadron collider, specifically for a proton collider corresponding to  the LHC  in Run 2. Studying the processes at LO, the relevant channels are the quark annihilation ones. At the level of the hard scattering, the kinematic dependence and therefore the density matrices are fairly similar to the corresponding ones at a lepton collider. The main differences are given by the different charges, which induce different couplings to the EW bosons.
We find these effects to be relevant but not particularly disruptive of the entanglement pattern. In particular, one can see that the explicit expressions for the density matrix in the benchmark regions analysed in the previous section are fairly similar.

What really changes the picture in a proton collider are the contributions from different partonic channels, weighted by the corresponding parton luminosity. Also, since we are now colliding identical particles, the system will be invariant under the transformation $\theta \to \pi + \theta$, showing therefore a symmetry with respect to $\cos{\theta}=0$. This can also be understood at the level of the $R$-matrix from the symmetrisation in \cref{eq:Rmat_pp}. As a general remark, the most evident effect of having to sum over the different partonic channels is that the presence of entanglement is considerably diluted.

\subsubsection*{$WW$ production}

\begin{figure}
    \centering
    \includegraphics[width=.9\linewidth]{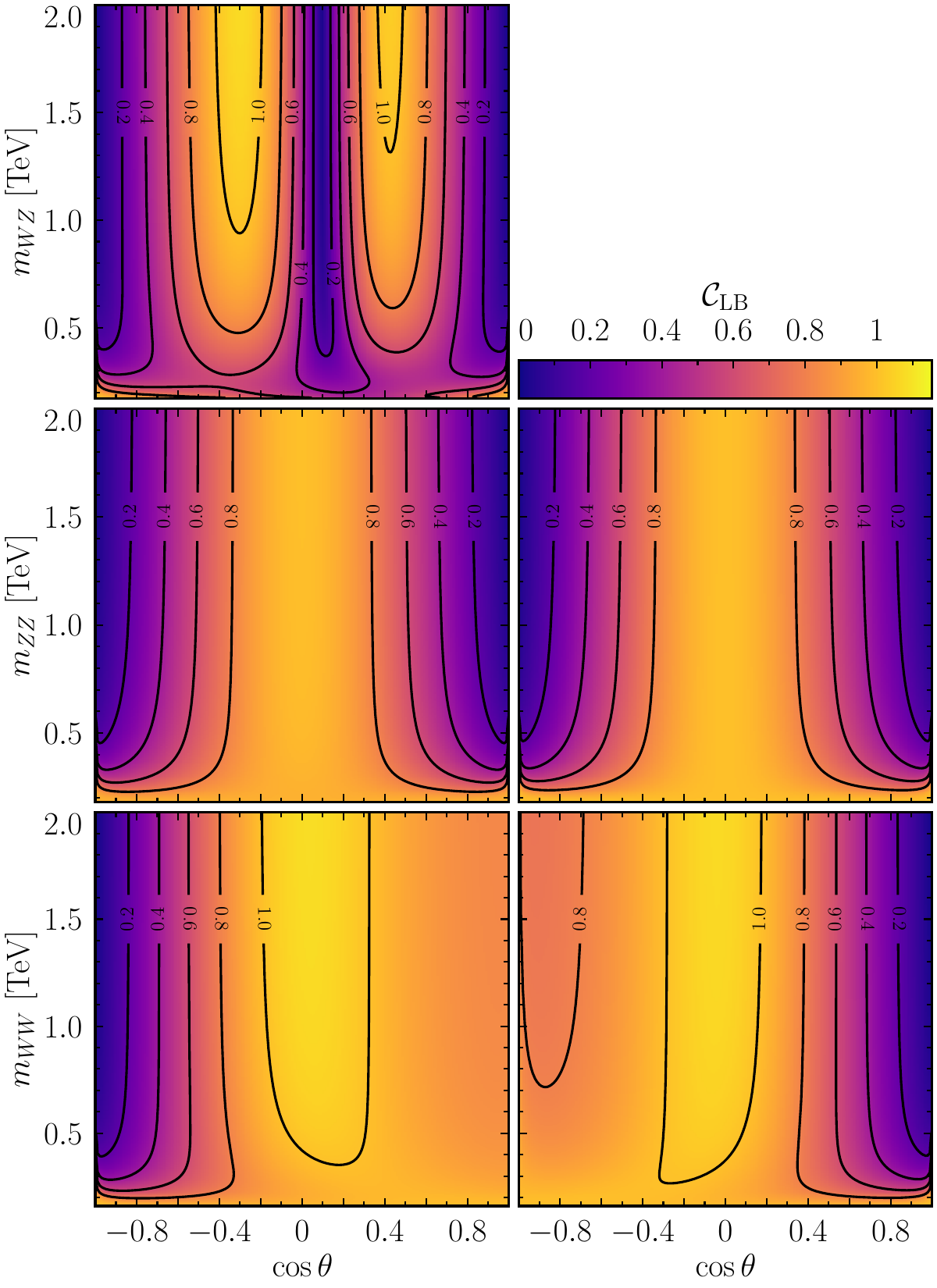}
    \caption{%
        The lower bound for the concurrence \cMB in the partonic channels as a function of the invariant mass $m_{VV}$ of the diboson pair and the cosine of the angle between the $\bar{q}$ and the $W^+$ (or $Z$ for $ZZ$), in the centre of mass frame, in the SM. 
        Bottom: $W^+ W^-$ production in the $u\bar{u}$~(left) and $d\bar{d}$~(right) channel.
        Middle: $Z Z$ production in the $u\bar{u}$~(left) and $d\bar{d}$~(right) channel.
        Top: $u\bar{d}\to Z W^+$ production.
    }
    \label{fig:entang_qq}
\end{figure}

\begin{figure}
    \centering
    \includegraphics[width=.9\linewidth]{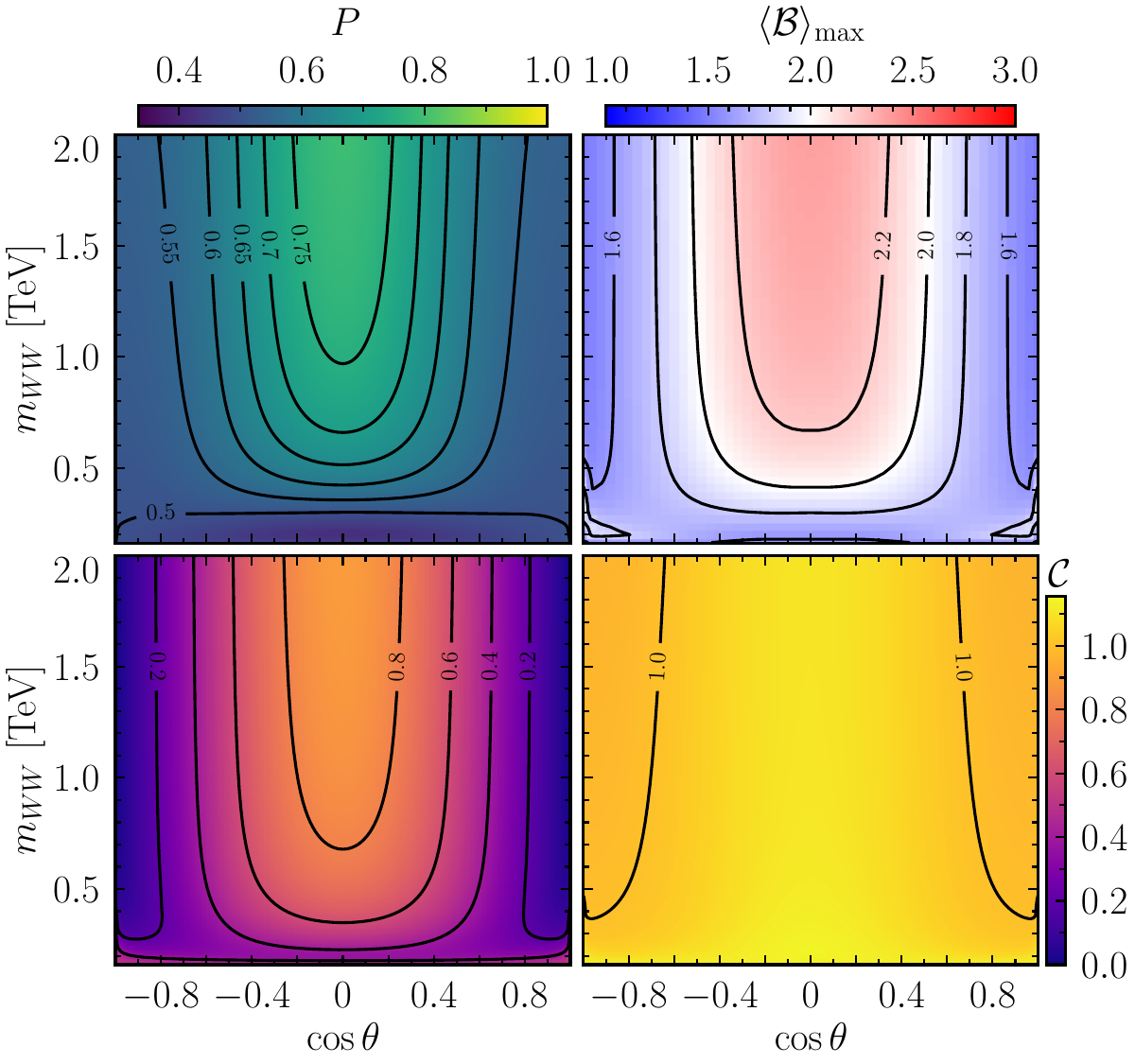}
    \caption{%
        Entanglement in the $pp \to W^+ W^-$ channel in the SM. We show the lower bound~\cMB~(bottom left) and upper bound~\cUB~(bottom right) on the concurrence \concurrence, the purity~$P$~(top left) as well as the indicator~$\langle\mathcal{B}\rangle_\mathrm{max}$ for Bell inequality violation~(top right) as a function of the invariant mass $m_{WW}$ and the cosine of the angle between the proton and the $W^+$, in the centre of mass frame.
    }
    \label{fig:entang_ppWW}
\end{figure}

As in the case of a lepton collider, the production of a pair of $W$ bosons is the dominant diboson production channel. The high cross section makes it an ideal candidate to probe the EW interactions and potentially uncover signs of NP in the tails of the distributions. With this aim,  the study of the spin density matrix can offer a complementary approach which could be helpful to disentangle degeneracies and characterise potential signals.

In the bottom panel of \cref{fig:entang_qq}, we show the value of the marker \cMB across phase space for the two independent channels in proton collisions, $u \, \bar{u} \to W^+ \, W^-$ and $d \, \bar{d} \to W^+ \, W^-$. As opposed to the lepton collider, we decide to show results up to \SI{2}{\TeV} in invariant mass. The physics regulating the production of $W$ bosons is fairly similar for all initial states and the only differences are given by the different couplings the particles have with the $Z$ boson, \ie the \mbox{$s$-channel} diagram.
As a consequence, the balance between the right-handed and left-handed couplings of the various initial states is slightly different and this translates into a slight difference for the density matrix. Note that the mirroring of the $u \bar{u}$ channel is fictitious and simply given by the conventional choice of the angle between the anti-up quark and the $W^+$. In fact, in the case of the positron and the $d$ quark, that angle is the one between the initial and final state particles that share the same-sign charge, while in the $u \bar{u}$ case it is the angle between the opposite charge particles.

In the lower panel of \cref{fig:entang_ppWW}, we plot the lower limit of the concurrence \cMB for the parton luminosity weighted combination of the channels, as described in \cref{eq:Rmat_pp}. As expected, the result is symmetric with respect to $\cos{\theta}=0$, given the symmetrisation over the polar angle. It is interesting to observe the strong dilution of the entanglement pattern, which is caused precisely by summing over the initial state and considering both $q\bar{q}$ and $\bar{q}q$ channels. This can be intuitively understood by looking at the plots of the individual channels, where we can observe that the two collinear regions, $\cos{\theta}=1$ and $\cos{\theta}=-1$ are characterised by opposite behaviours and therefore the high entanglement of one region is washed out by the other when summing over the two different polar angles in \cref{eq:Rmat_pp}. Because of this, the diboson pair produced will mostly be in a mixed state, and a high level of entanglement will only be found in the central region and at high energy, around $\theta = \pi/2$.

As for the lepton collider case, in \cref{fig:entang_ppWW} we also report on the value of the quantum observable indicators for the purity $P$, the Bell inequality violation marker $\langle\mathcal{B}\rangle_\mathrm{max}$ and the upper bound on the concurrence \cUB. The latter is not bringing much information to the table, since it shows values of order 1 across phase space. On the other hand, from the purity plot we learn that the density matrix is characterised by a highly mixed state and it goes towards a purer state in the high-energy central region. We find that this is a common feature across all proton collider processes, mostly an effect of the state mixing dictated by \cref{eq:Rmat_pp}. Finally, in the upper right plot of \cref{fig:entang_ppWW} we display the marker for Bell inequality violation, which follows closely the pattern indicated by the marker \cMB, sign once again that states violating Bell inequalities are a subset of entangled states.
\begin{figure}[t!]
    \centering
    \includegraphics[width=.9\linewidth]{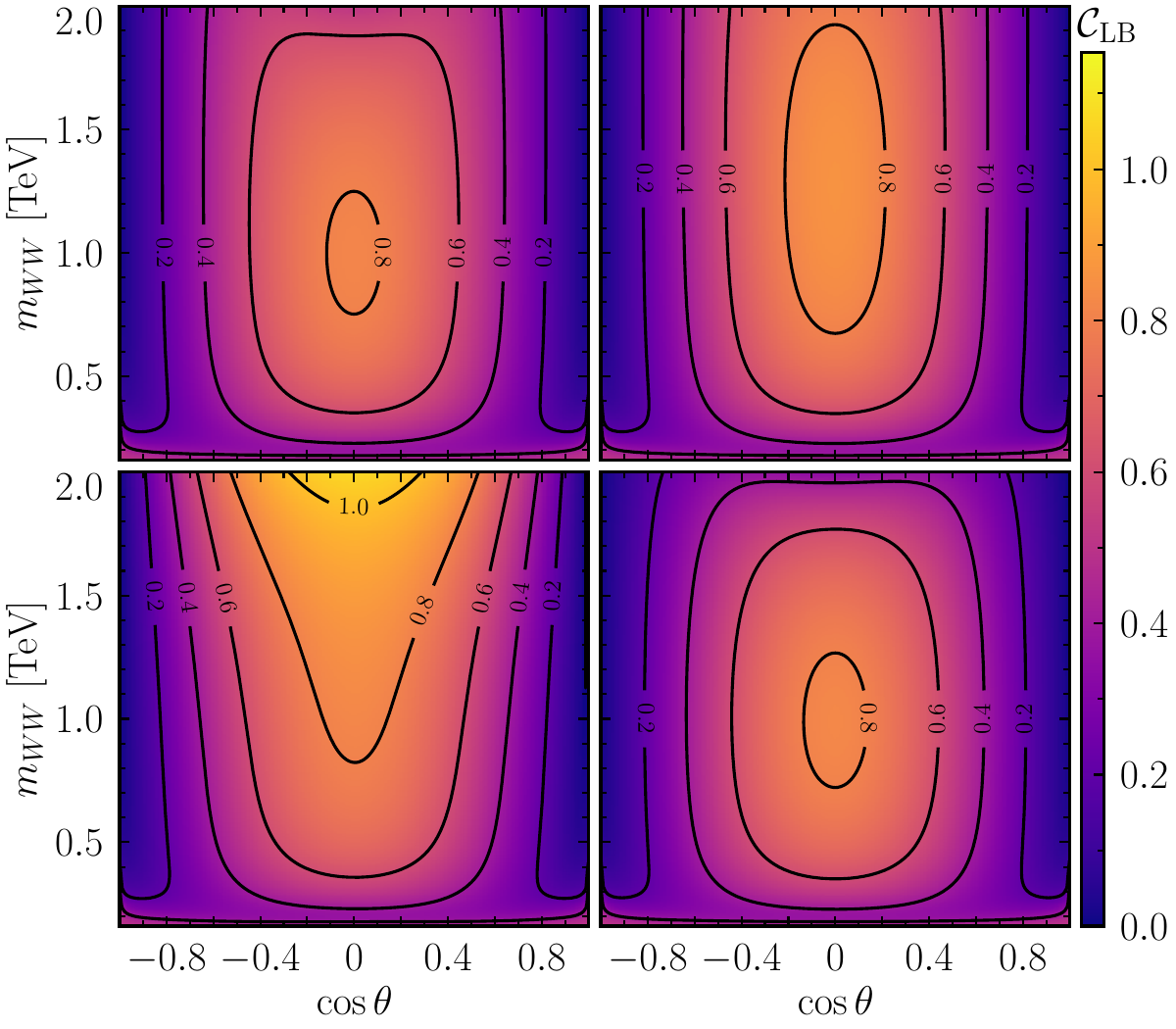}
    \caption{The changes in the marker \cMB is shown for a selection of operators and benchmark Wilson coefficient values for $W^+W^-$ production at a proton collider. Only one operator at the time is switched on. Top left: $c_{\varphi u} = \SI{0.05}{\TeV^{-2}}$, top right: $c_{\varphi d} = \SI{0.05}{\TeV^{-2}}$, bottom left: $c_{\varphi q}^{(3)} = \SI{0.05}{\TeV^{-2}}$, bottom right: $c_{W} = \SI{0.03}{\TeV^{-2}}$.}
    \label{fig:entang_ppWW_EFT}
\end{figure}

\begin{figure}[t!]
    \centering
    \includegraphics[width=\linewidth]{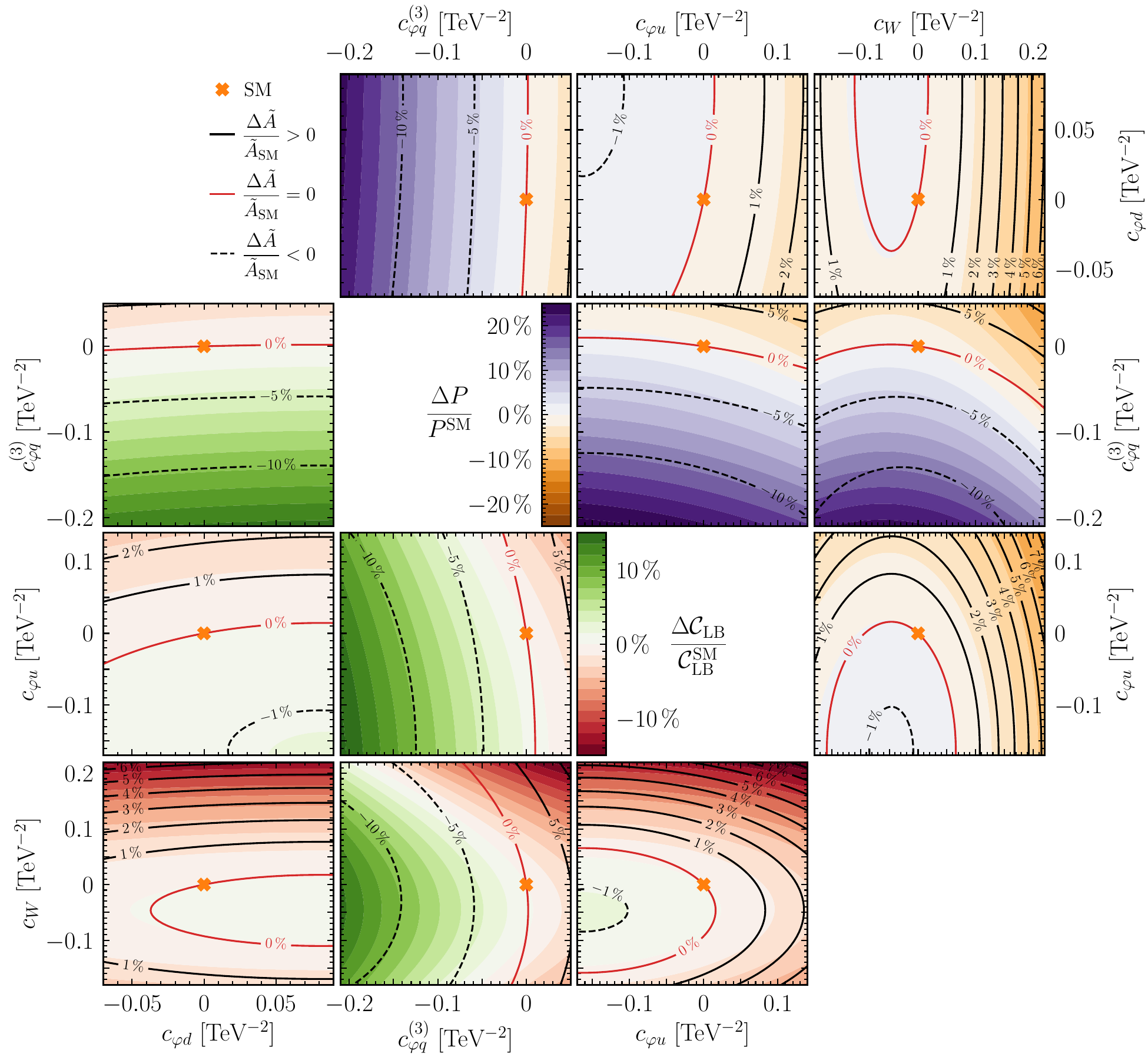}
    \caption{The relative changes in the marker \cMB~(lower triangle) and purity~$P$~(upper triangle) compared to the SM values $\cMB^\mathrm{SM} = 0.73$ and $P^\mathrm{SM}=0.64$ as a function of the Wilson coefficients $c_{\varphi d}$, $c_{\varphi q}^{(3)}$, $c_{\varphi u}$ and $c_{W}$ for $W^+W^-$ production at a proton collider, at $m_{WW}=\SI{500}{\GeV}$ and $\cos\theta=0$.
        The lines indicate the relative change of the cross section.}
        \label{fig:entang_ppWW_EFT_triangle}
\end{figure}

We now move on to the study of the EFT effects to the density matrix of the $W$ pair produced in a proton collider. In \cref{fig:entang_ppWW_EFT} we display the changes in the \cMB marker for some benchmark Wilson coefficients. As can be seen in the two top figures, the effects of $c_{\varphi u}$ and $c_{\varphi d}$ are very similar, enhancing the right-handed coupling of the $Z$ boson with the quarks and consequently decreasing the level of entanglement at high energy. The different intensity of the two operators has to be traced back to the different weight at the level of the PDFs, \ie, the $u\bar{u}$ luminosity is higher than the $d\bar{d}$ one. As expected, switching on the $c_{\varphi q}^{(3)}$, which instead modifies the left-handed coupling to the $Z$ of the $d$-quarks and the coupling to the $W$ boson of both $u$ and $d$, has the opposite effect, increasing the value of the concurrence in the central region where we see the emergence of maximal level of entanglement. A similar effect is found for the $c_{\varphi q}^{-}$ Wilson coefficient (not displayed) which modifies the left-handed coupling of the $u$ quarks with the $Z$ boson. Finally, in the lower-right plot in \cref{fig:entang_ppWW_EFT} we show the effects of the $\OO_{W}$ operator. Interestingly, we find that the presence of the pure BSM coupling $\delta \lambda_V$ can be quite disruptive, especially at high energy, inducing a decrease of the level of entanglement. The effects of the $c_{\varphi D}$ and $c_{\varphi W B}$, which modify the SM TGCs, are found to be sensibly smaller in this case.
We observe that the EFT effects are mostly in the central region while the collinear regions keep being characterised by a substantial absence of entanglement.

Finally, in \cref{fig:entang_ppWW_EFT_triangle}, we depict the relative change of \cMB~(lower triangle) and the purity~(upper triangle) as a function of the Wilson coefficients, varying two coefficients at a time and considering the fixed phase space point $m_{WW}=\SI{500}{\GeV}$ and $\theta=\pi/2$. 
Contrary to the case of a lepton collider, we do not find that we gain much from the chosen spin-related observables, \ie, the probed directions in parameter space are very similar to the ones probed by the differential cross section. This has to be traced back to the fact that in a proton collider we sum over the different partonic channels and in doing so we lose sensitivity to the spin related Fano coefficients. The quantum spin observables are indeed highly dependent on the Fano coefficient $\tilde{A}$, which controls the abundance of a channel over the other, and therefore affects the spin density matrix of the total system. We verified indeed that if one were to single out a specific channel (for example only $u \bar{u}$ and its $\bar{u} u$ counterpart), one would find results much more similar to those observed in \cref{fig:entang_eeWW_EFT_triangle} for the case of a lepton collider.

\subsubsection*{$ZZ$ production}

\begin{figure}
    \centering
    \includegraphics[width=.9\linewidth]{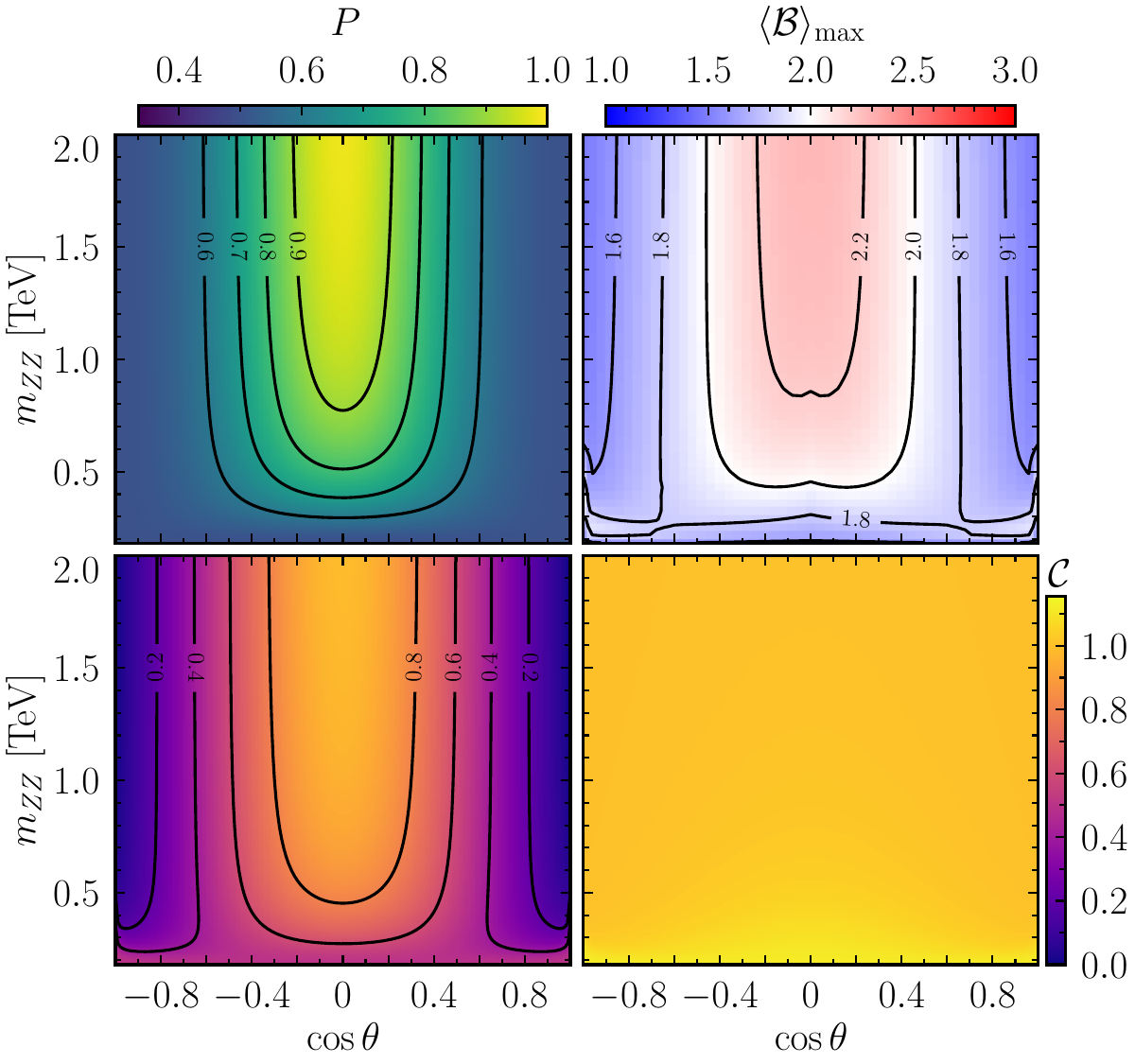}
    \caption{%
        Entanglement in the $pp \to Z Z$ channel in the SM. We show the lower bound~\cMB~(bottom left) and upper bound~\cUB~(bottom right) on the concurrence \concurrence, the purity~$P$~(top left) as well as the indicator~$\langle\mathcal{B}\rangle_\mathrm{max}$ for Bell inequality violation~(top right) as a function of the invariant mass $m_{ZZ}$ and the cosine of the angle between the proton and the $Z$, in the centre of mass frame.
    }
    \label{fig:entang_ppZZ}
\end{figure}

We now discuss $ZZ$ production at a proton collider. In the middle panel of \cref{fig:entang_qq}, we show the entanglement pattern for the case of quark annihilation and in \cref{fig:entang_ppZZ} their combination in a proton collider (lower left plot). It is interesting to notice that the quark channels present a different pattern of entanglement with respect to the lepton collider, which is caused by the simple fact that the value of the $\bar{g}_V$ coupling depends on the charge of the particle and is sensibly different in the three cases, \ie $\bar{g}_V \approx -0.027, 0.1, -0.17$ for $e$, $u$ and $d$ respectively. This is non-trivial as one could have naively expected the patterns to be the same given that the EW couplings involved are the same. Also in the case of a proton collider, the general feature remains that high entanglement in $Z Z$ production can be found at high invariant mass and in the central region. Once again the plot of the upper bound marker \cUB does not deliver any information, as the indicator reaches close to maximal values everywhere in phase space. The plot of the Bell violating marker $\langle\mathcal{B}\rangle_\mathrm{max}$ in the upper right corner of \cref{fig:entang_ppZZ} confirms that the density matrix in the central high-energy region is characterised by highly entangled quantum states.

Moving on to the EFT effects, as we already discussed in the corresponding section on $ZZ$ production at a lepton collider, the only possible modifications are given by the shift of the vectorial and axial couplings to the $Z$ boson, in particular from operators that spoil the balance between the two. We already saw that the effects are more subtle compared to the case of $W^+ W^-$ production. However, rather surprisingly we observe that in the case of the proton collisions there is even more sturdiness towards EFT effects. For values of the Wilson coefficients within the current bounds coming from global fits, no visible effect on the pattern of entanglement is present. For this reason we believe that $Z$ pair production is the least promising process to probe dimension-6 effects at the level of the spin density matrix.

\subsubsection*{$WZ$ production}

\begin{figure}
    \centering
    \includegraphics[width=.9\linewidth]{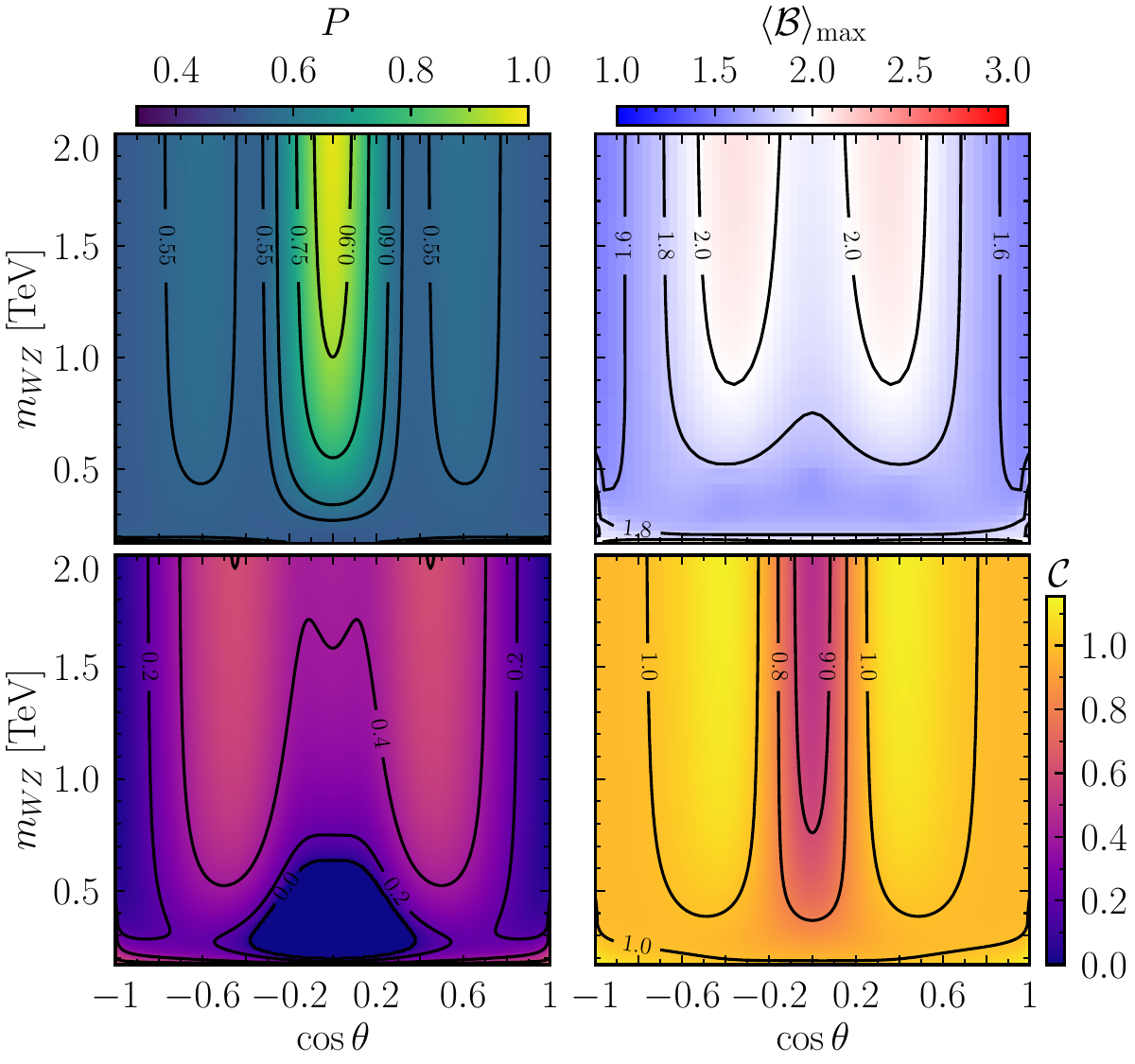}
    \caption{%
        Entanglement in the $pp \to W^+ Z$ channel in the SM. We show the lower bound~\cMB~(bottom left) and upper bound~\cUB~(bottom right) on the concurrence \concurrence, the purity~$P$~(top left) as well as the indicator~$\langle\mathcal{B}\rangle_\mathrm{max}$ for Bell inequality violation~(top right) as a function of the invariant mass $m_{WZ}$ and the cosine of the angle between the proton and the $W^+$, in the centre of mass frame.
    }
    \label{fig:entang_ppWZ}
\end{figure}

Contrary to the previous diboson production modes, $W Z$ cannot be produced at a $e^+ e^-$ collider without the emission of additional charged particles.
At a proton collider, only one relevant partonic channel exists, \ie, $u \bar{d} \to W^+ Z$ and the charge conjugated one. In the following we will focus on $W^+ Z$ production as representative of the two processes, which can distinguished in experiments. In this process, the relevant couplings are those of the  $W$ and $Z$ to the fermions, as well as the triple gauge coupling. Note that in the case of $W Z$, at the partonic level depicted in the top panel of \cref{fig:entang_qq}, the expressions for \cMB and \cUB are identical (in the SM), indicating that \cMB is precisely the value of the concurrence and not just a lower bound. This is ultimately due to the fact that the trace of $\rho^2$ is equal to 1 (the system is a pure state everywhere in phase space) and therefore the expressions in \cref{eq:cmb,eq:cub} coincide. 
The statement does not necessarily hold true anymore in the presence of modified interactions, as we verified for the dimension-6 operators considered in this work. However, this property is not maintained once the symmetrisation of the $\theta$ angle in \cref{eq:Rmat_pp} is performed and consequently the density matrix for proton collisions is ultimately highly mixed, as can be seen in the upper left plot in \cref{fig:entang_ppWZ}. To improve the purity, one could consider events boosted in the forward/backward regions and try to infer on a statistical basis the directions of the quark and anti-quark in the initial states.

In \cref{fig:entang_qq} (top plot) we show the \cMB pattern for the individual channel, while the proton collider one is displayed in the bottom left plot in \cref{fig:entang_ppWZ}, where the main difference is due to the fact that we have to take into account both $q\bar{q}$ and $\bar{q} q$ initial states, cf.\ \cref{eq:Rmat_pp}. We notice once again, that the entanglement is much lower in the proton collider with respect to the individual channel. Surprisingly, we find high entanglement at threshold as well as for $\cos{\theta} \approx \pm 0.5$, while in the central region, for $\cos{\theta} = 0$, the value of the concurrence is low. The overall pattern differs substantially with respect to the previously analysed diboson processes. As a matter of fact, as can be seen from the marker $\langle\mathcal{B}\rangle_\mathrm{max}$ the violation of Bell inequalities is really weak even at high energy, mostly a consequence of the fact that the density matrix is in a high mixture of states (see upper left plot in \cref{fig:entang_ppWZ}).

With respect to the quantum state produced by the $u \bar{d}$ channel, we find that at threshold, the density matrix is described by a pure state
\begin{equation}
    \ket{\Psi(m_{WZ} = m_W + m_Z)} \approx 0.75 \ket{0+}_{\boldsymbol{p}} + 0.66 \ket{+0}_{\boldsymbol{p}} \, .
\end{equation}
Notably, the quantum state is not symmetric under label exchange, as a consequence of the fact that we are not dealing any more with pairs of particles sharing the same mass and interactions.  At high energy, in the collinear limit $\theta = 0$, the density matrix is described by a pure separable state, \ie, $\ket{++}_{\boldsymbol{p}}$.
On the other hand, in the central region, we still have a pure but partially entangled state
\begin{equation}
    \ket{\Psi(m_{WZ} \to \infty, \cos{\theta}=0)} \approx 0.164 \ket{++}_{\boldsymbol{k}} - 0.973 \ket{00}_{\boldsymbol{k}} - 0.164 \ket{- -}_{\boldsymbol{k}} \, .
\end{equation}
\begin{figure}[t!]
    \centering
    \includegraphics[width=\linewidth]{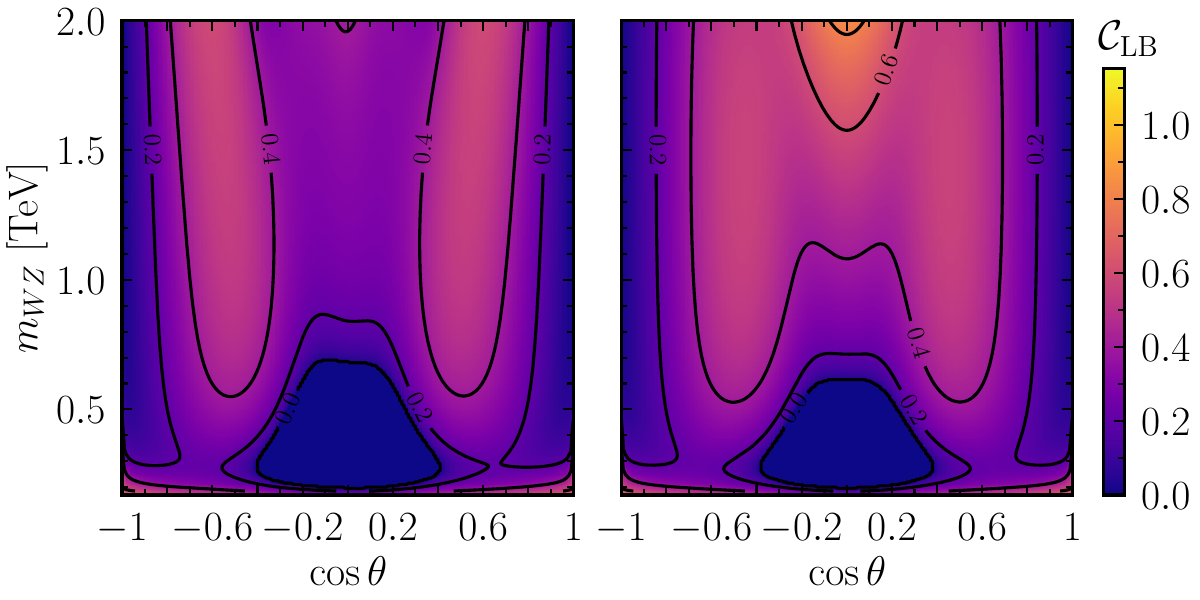}
    \caption{The changes in the marker \cMB is shown for a selection of operators and benchmark Wilson coefficient values for $W^+ Z$ production at a proton collider. Only one operator at the time is switched on. Left: $c_{\varphi q}^{(3)} = \SI{0.03}{\TeV^{-2}}$, Right: $c_{W} = \SI{0.01}{\TeV^{-2}}$.}
    \label{fig:entang_ppWZ_EFT}
\end{figure}
\begin{figure}[t!]
    \centering
    \includegraphics[width=.9\linewidth]{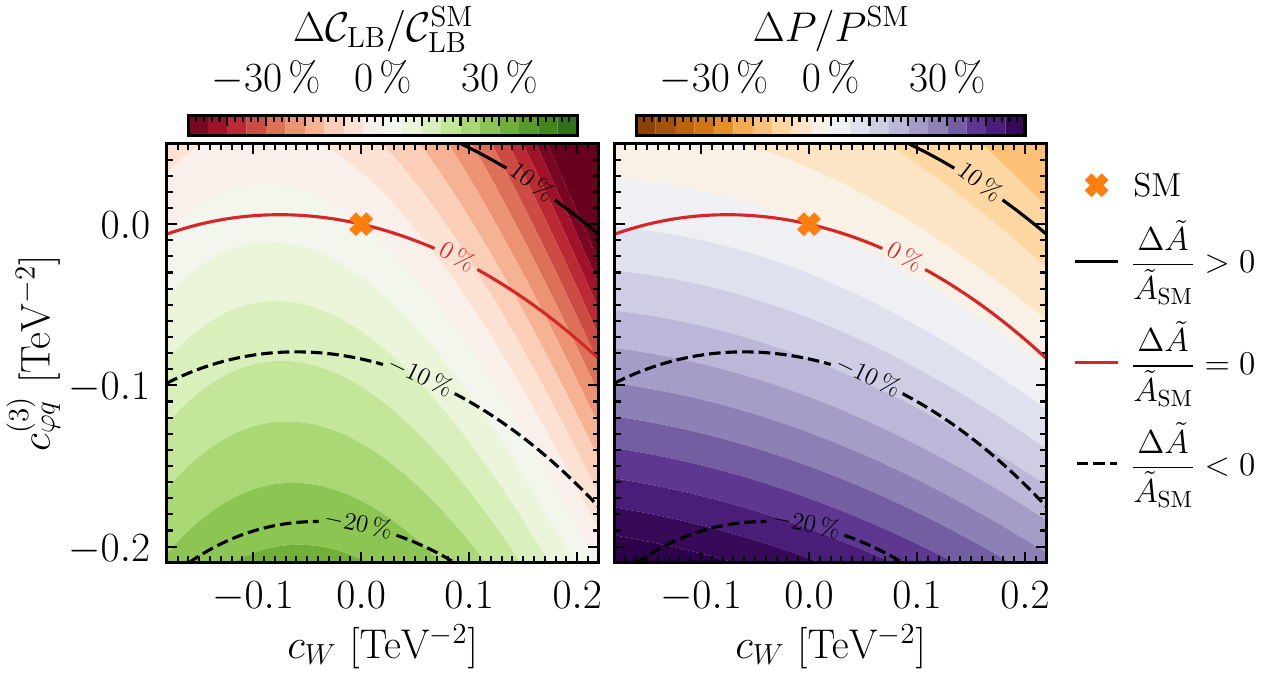}
    \caption{The relative changes in the marker \cMB~(lower triangle) and purity~$P$~(upper triangle) compared to the SM values $\cMB^\mathrm{SM} = 0.38$ and $P^\mathrm{SM}=0.55$ as a function of the Wilson coefficients $c_{W}$ and $c_{\varphi q}^{(3)}$ for $W Z$ production at a proton collider, at $M_{WZ}=\SI{500}{\GeV}$ and $\cos\theta=1/2$.
        The lines indicate the relative change of the cross section.}
           \label{fig:entang_ppWZ_EFT_triangle}
\end{figure}
\noindent Finally, we discuss the effects of the dimension-6 EFT operators. We point out that, for this specific process, we set $m_W=m_Z$ when computing the EFT corrections. This approximation considerably speeds up the computation, and given that higher dimensional operator corrections are mostly relevant at high energy, we expect the difference $m_Z-m_W$ to be negligible. We have verified the former statement for a subset of the operators, finding that the naive expectation holds true. 

In \cref{fig:entang_ppWZ_EFT} we display the changes in the \cMB marker for some benchmark Wilson coefficients. We find that the density matrix is particularly sensitive to higher dimensional operators for this process, even within the current bounds set by global EFT fits. In particular in the plot on the right, we observe a strong effect coming from the presence of the Wilson coefficient $c_{W}$, which is unique in its kind inducing the BSM coupling $\delta \lambda_V$. The effect of this operator is to increase the entanglement in the central region at high energy, indicating that the dominant configuration in that phase space region is enhanced by the presence of the operator. As previously discussed, the $\OO_W$ operator is of particular interest from an EFT perspective and we find that the $W^+ Z$ final state seems to be the best probe among the diboson final states with respect to the spin density matrix.
In the plot on the left in \cref{fig:entang_ppWZ_EFT} we instead show the changes coming from a modification of the left-handed coupling to the EW bosons by the $\OO_{\varphi q}^{(3)}$ operator. The effect in this case is opposite and the value of the \cMB marker is generally decreased in the high energy phase space region, indicating that the two operators enhance different spin configurations.
We find that other operators affecting the process ($\OO_{\varphi q}^{(1)}$, $\OO_{\varphi D}$ and $\OO_{\varphi WB}$) have close to negligible effects on the entanglement pattern when considering non-excluded values of the Wilson coefficients. We therefore do not show the corresponding plots.

To conclude the section, in \cref{fig:entang_ppWZ_EFT_triangle} we display the relative change of \cMB~(left) and the purity~(right) as a function of the Wilson coefficients in two-dimensional planes. Note that contrary to previously discussed processes, here the phase space point chosen is different as in the central phase space region $\theta=\pi/2$, low values of entanglement are found and specifically for $m_{WZ}=\SI{500}{\GeV}$, the value of \cMB is equal to $0$. We therefore opt for showing plots for $\cos{\theta}=1/2$. Once again, the added value of the spin observables in searches for new heavy physics is evident and the plot displays a significant sensitivity both at the cross section level and at the level of the spin density matrix, with deviations from the SM predictions as big as \SI{30}{\percent} for values of the Wilson coefficients well within the current limits from global fits.

\section{Conclusions}
\label{sec:Outro}

In this work, we have explored the sensitivity of quantum observables to  the strength and structure of couplings entering diboson production, both in the context of the SM and of the SMEFT. Our main objective  was to gauge the power of the quantum spin density matrix and related observables to probe the existence of NP.

After setting up the formalism, we studied the behaviour of scattering amplitudes in the high energy regime, where longitudinal polarisations dominate the production mode, assessing the consequences for the spin density matrix. In particular, we exploited the known fact that spin observables allow to study effects that are subdominant at the level of single particle distributions, and depend on the interference between higher-dimensional operators and the SM. These observables display a sensitivity to deviations from the SM predictions in the high-energy tails of the distributions, which will be further explored with Run-3 and the HL-LHC in the coming decade. 

The main results of our study are reported in \cref{sec:Entanglement_colliders}, where different processes have been analysed. We have considered both lepton and proton colliders, finding that the former offer a much cleaner setup for spin density matrix probes. This is mostly due to the fact that in a proton collider the quantum state of the system is the incoherent sum of different partonic channels and therefore tends to be mixed.  Nonetheless, considerable sensitivity to NP is also found at proton colliders, which,  featuring higher centre of mass energies, can take full advantage of the energy growth of the dimension-6 amplitudes. 

In general, we find that the $ZZ$ production is the least interesting process when it comes to NP sensitivity, as the phenomenology is completely determined by only two possibly anomalous couplings (the right-handed and the left-handed coupling to the $Z$ boson) and the dimension-6 operators do not introduce new Lorentz structures. We note, however, the potential interest in studying the effects of the neutral TGC which arise at dimension-8. In this case, spin-observables could help in gaining sensitivity, especially because of the possibility to fully reconstruct the final state, something which is experimentally more challenging for final states involving $W$ bosons.  On the other hand, we find that $WW$ and $WZ$ production show a rather large sensitivity to heavy NP effects in the spin density matrix already at dimension-6 with significant changes expected in the entanglement pattern across phase space.  For example,  interference effects due to the triple gauge operator $\mathcal{O}_W$ are clearly identified by quantum observables. 

Our results motivate an experimental feasibility study for performing the detailed quantum tomography of the four-fermion final states arising from $VV$ production in the SMEFT framework at the LHC and at future lepton colliders.

\section*{Acknowledgements}

RA's research was supported by the F.R.S.-FNRS project no.~40005600 and the FSR Program of UCLouvain.
FM is partially supported by the F.R.S.-
FNRS under the “Excellence of Science” EOS be.h project no. 30820817. 
LM is supported by the European Research Council under the European Union’s Horizon 2020 research and innovation Programme (grant agreement n.950246).

\bibliographystyle{JHEP}
\bibliography{references}

\providecommand{\href}[2]{#2}\begingroup\raggedright\begin{thebibliography}{10}

\bibitem{Cervera-Lierta:2017tdt}
A.~Cervera-Lierta, J.I.~Latorre, J.~Rojo and L.~Rottoli, \emph{{Maximal
  Entanglement in High Energy Physics}},
  \href{https://doi.org/10.21468/SciPostPhys.3.5.036}{\emph{SciPost Phys.}
  {\bfseries 3} (2017) 036} [\href{https://arxiv.org/abs/1703.02989}{{\ttfamily
  1703.02989}}].

\bibitem{Beane:2018oxh}
S.R.~Beane, D.B.~Kaplan, N.~Klco and M.J.~Savage, \emph{{Entanglement
  Suppression and Emergent Symmetries of Strong Interactions}},
  \href{https://doi.org/10.1103/PhysRevLett.122.102001}{\emph{Phys. Rev. Lett.}
  {\bfseries 122} (2019) 102001}
  [\href{https://arxiv.org/abs/1812.03138}{{\ttfamily 1812.03138}}].

\bibitem{Low:2021ufv}
I.~Low and T.~Mehen, \emph{{Symmetry from entanglement suppression}},
  \href{https://doi.org/10.1103/PhysRevD.104.074014}{\emph{Phys. Rev. D}
  {\bfseries 104} (2021) 074014}
  [\href{https://arxiv.org/abs/2104.10835}{{\ttfamily 2104.10835}}].

\bibitem{Afik:2020onf}
Y.~Afik and J.R.M.~de~Nova, \emph{{Entanglement and quantum tomography with top
  quarks at the LHC}},
  \href{https://doi.org/10.1140/epjp/s13360-021-01902-1}{\emph{Eur. Phys. J.
  Plus} {\bfseries 136} (2021) 907}
  [\href{https://arxiv.org/abs/2003.02280}{{\ttfamily 2003.02280}}].

\bibitem{Fabbrichesi:2021npl}
M.~Fabbrichesi, R.~Floreanini and G.~Panizzo, \emph{{Testing Bell Inequalities
  at the LHC with Top-Quark Pairs}},
  \href{https://doi.org/10.1103/PhysRevLett.127.161801}{\emph{Phys. Rev. Lett.}
  {\bfseries 127} (2021) 161801}
  [\href{https://arxiv.org/abs/2102.11883}{{\ttfamily 2102.11883}}].

\bibitem{Severi:2021cnj}
C.~Severi, C.D.E.~Boschi, F.~Maltoni and M.~Sioli, \emph{{Quantum tops at the
  LHC: from entanglement to Bell inequalities}},
  \href{https://doi.org/10.1140/epjc/s10052-022-10245-9}{\emph{Eur. Phys. J. C}
  {\bfseries 82} (2022) 285}
  [\href{https://arxiv.org/abs/2110.10112}{{\ttfamily 2110.10112}}].

\bibitem{Afik:2022kwm}
Y.~Afik and J.R.M.~de~Nova, \emph{{Quantum information with top quarks in
  QCD}}, \href{https://doi.org/10.22331/q-2022-09-29-820}{\emph{Quantum}
  {\bfseries 6} (2022) 820} [\href{https://arxiv.org/abs/2203.05582}{{\ttfamily
  2203.05582}}].

\bibitem{Aguilar-Saavedra:2022uye}
J.A.~Aguilar-Saavedra and J.A.~Casas, \emph{{Improved tests of entanglement and
  Bell inequalities with LHC tops}},
  \href{https://doi.org/10.1140/epjc/s10052-022-10630-4}{\emph{Eur. Phys. J. C}
  {\bfseries 82} (2022) 666}
  [\href{https://arxiv.org/abs/2205.00542}{{\ttfamily 2205.00542}}].

\bibitem{Dong:2023xiw}
Z.~Dong, D.~Gon\c{c}alves, K.~Kong and A.~Navarro, \emph{{When the Machine
  Chimes the Bell: Entanglement and Bell Inequalities with Boosted
  $t\bar{t}$}},  \href{https://arxiv.org/abs/2305.07075}{{\ttfamily
  2305.07075}}.

\bibitem{Afik:2022dgh}
Y.~Afik and J.R.M.~de~Nova, \emph{{Quantum discord and steering in top quarks
  at the LHC}},  \href{https://arxiv.org/abs/2209.03969}{{\ttfamily
  2209.03969}}.

\bibitem{Aoude:2022imd}
R.~Aoude, E.~Madge, F.~Maltoni and L.~Mantani, \emph{{Quantum SMEFT tomography:
  Top quark pair production at the LHC}},
  \href{https://doi.org/10.1103/PhysRevD.106.055007}{\emph{Phys. Rev. D}
  {\bfseries 106} (2022) 055007}
  [\href{https://arxiv.org/abs/2203.05619}{{\ttfamily 2203.05619}}].

\bibitem{Severi:2022qjy}
C.~Severi and E.~Vryonidou, \emph{{Quantum entanglement and top spin
  correlations in SMEFT at higher orders}},
  \href{https://arxiv.org/abs/2210.09330}{{\ttfamily 2210.09330}}.

\bibitem{Fabbrichesi:2022ovb}
M.~Fabbrichesi, R.~Floreanini and E.~Gabrielli, \emph{{Constraining new physics
  in entangled two-qubit systems: top-quark, tau-lepton and photon pairs}},
  \href{https://arxiv.org/abs/2208.11723}{{\ttfamily 2208.11723}}.

\bibitem{Altakach:2022ywa}
M.M.~Altakach, P.~Lamba, F.~Maltoni, K.~Mawatari and K.~Sakurai, \emph{{Quantum
  information and CP measurement in
  H\textrightarrow{}\ensuremath{\tau}+\ensuremath{\tau}- at future lepton
  colliders}}, \href{https://doi.org/10.1103/PhysRevD.107.093002}{\emph{Phys.
  Rev. D} {\bfseries 107} (2023) 093002}
  [\href{https://arxiv.org/abs/2211.10513}{{\ttfamily 2211.10513}}].

\bibitem{Barr:2021zcp}
A.J.~Barr, \emph{{Testing Bell inequalities in Higgs boson decays}},
  \href{https://doi.org/10.1016/j.physletb.2021.136866}{\emph{Phys. Lett. B}
  {\bfseries 825} (2022) 136866}
  [\href{https://arxiv.org/abs/2106.01377}{{\ttfamily 2106.01377}}].

\bibitem{Barr:2022wyq}
A.J.~Barr, P.~Caban and J.~Rembieli\'nski, \emph{{Bell-type inequalities for
  systems of relativistic vector bosons}},
  \href{https://arxiv.org/abs/2204.11063}{{\ttfamily 2204.11063}}.

\bibitem{Ashby-Pickering:2022umy}
R.~Ashby-Pickering, A.J.~Barr and A.~Wierzchucka, \emph{{Quantum state
  tomography, entanglement detection and Bell violation prospects in weak
  decays of massive particles}},
  \href{https://arxiv.org/abs/2209.13990}{{\ttfamily 2209.13990}}.

\bibitem{Fabbrichesi:2023cev}
M.~Fabbrichesi, R.~Floreanini, E.~Gabrielli and L.~Marzola, \emph{{Bell
  inequalities and quantum entanglement in weak gauge bosons production at the
  LHC and future colliders}},
  \href{https://arxiv.org/abs/2302.00683}{{\ttfamily 2302.00683}}.

\bibitem{Aguilar-Saavedra:2022mpg}
J.A.~Aguilar-Saavedra, \emph{{Laboratory-frame tests of quantum entanglement in
  $H \to WW$}},  \href{https://arxiv.org/abs/2209.14033}{{\ttfamily
  2209.14033}}.

\bibitem{Aguilar-Saavedra:2022wam}
J.A.~Aguilar-Saavedra, A.~Bernal, J.A.~Casas and J.M.~Moreno, \emph{{Testing
  entanglement and Bell inequalities in $H \to ZZ$}},
  \href{https://arxiv.org/abs/2209.13441}{{\ttfamily 2209.13441}}.

\bibitem{Fabbrichesi:2023jep}
M.~Fabbrichesi, R.~Floreanini, E.~Gabrielli and L.~Marzola, \emph{{Stringent
  bounds on $HWW$ and $HZZ$ anomalous couplings with quantum tomography at the
  LHC}},  \href{https://arxiv.org/abs/2304.02403}{{\ttfamily 2304.02403}}.

\bibitem{Morales:2023gow}
R.A.~Morales, \emph{{Exploring Bell inequalities and quantum entanglement in
  vector boson scattering}},
  \href{https://arxiv.org/abs/2306.17247}{{\ttfamily 2306.17247}}.

\bibitem{Aguilar-Saavedra:2023hss}
J.A.~Aguilar-Saavedra, \emph{{Post-decay quantum entanglement in top pair
  production}},  \href{https://arxiv.org/abs/2307.06991}{{\ttfamily
  2307.06991}}.

\bibitem{Hill:1997pfa}
S.~Hill and W.K.~Wootters, \emph{{Entanglement of a pair of quantum bits}},
  \href{https://doi.org/10.1103/PhysRevLett.78.5022}{\emph{Phys. Rev. Lett.}
  {\bfseries 78} (1997) 5022}
  [\href{https://arxiv.org/abs/quant-ph/9703041}{{\ttfamily
  quant-ph/9703041}}].

\bibitem{10.1063/1.2795840}
R.~Hildebrand, \emph{{Concurrence revisited}},
  \href{https://doi.org/10.1063/1.2795840}{\emph{Journal of Mathematical
  Physics} {\bfseries 48} (2007) 102108}.

\bibitem{PhysRevLett.98.140505}
F.~Mintert and A.~Buchleitner, \emph{Observable entanglement measure for mixed
  quantum states},
  \href{https://doi.org/10.1103/PhysRevLett.98.140505}{\emph{Phys. Rev. Lett.}
  {\bfseries 98} (2007) 140505}.

\bibitem{Zhang_2008}
C.-J.~Zhang, Y.-X.~Gong, Y.-S.~Zhang and G.-C.~Guo, \emph{Observable estimation
  of entanglement for arbitrary finite-dimensional mixed states},
  \href{https://doi.org/10.1103/physreva.78.042308}{\emph{Physical Review A}
  {\bfseries 78} (2008) }.

\bibitem{Uhlmann:1996mk}
A.~Uhlmann, \emph{{Optimizing entropy relative to a channel or a subalgebra}},
  in \emph{{21st International Colloquium on Group Theoretical Methods in
  Physics}}, 7, 1996 [\href{https://arxiv.org/abs/quant-ph/9701014}{{\ttfamily
  quant-ph/9701014}}].

\bibitem{Horodecki:2009zz}
R.~Horodecki, P.~Horodecki, M.~Horodecki and K.~Horodecki, \emph{{Quantum
  entanglement}}, \href{https://doi.org/10.1103/RevModPhys.81.865}{\emph{Rev.
  Mod. Phys.} {\bfseries 81} (2009) 865}
  [\href{https://arxiv.org/abs/quant-ph/0702225}{{\ttfamily
  quant-ph/0702225}}].

\bibitem{PhysRevA.78.042308}
C.-J.~Zhang, Y.-X.~Gong, Y.-S.~Zhang and G.-C.~Guo, \emph{Observable estimation
  of entanglement for arbitrary finite-dimensional mixed states},
  \href{https://doi.org/10.1103/PhysRevA.78.042308}{\emph{Phys. Rev. A}
  {\bfseries 78} (2008) 042308}.

\bibitem{Bell:1964kc}
J.S.~Bell, \emph{{On the Einstein-Podolsky-Rosen paradox}},
  \href{https://doi.org/10.1103/PhysicsPhysiqueFizika.1.195}{\emph{Physics
  Physique Fizika} {\bfseries 1} (1964) 195}.

\bibitem{Clauser:1969ny}
J.F.~Clauser, M.A.~Horne, A.~Shimony and R.A.~Holt, \emph{{Proposed experiment
  to test local hidden variable theories}},
  \href{https://doi.org/10.1103/PhysRevLett.23.880}{\emph{Phys. Rev. Lett.}
  {\bfseries 23} (1969) 880}.

\bibitem{PhysRevA.65.032118}
D.~Kaszlikowski, L.C.~Kwek, J.-L.~Chen, M.~\ifmmode~\dot{Z}\else
  \.{Z}\fi{}ukowski and C.H.~Oh, \emph{Clauser-horne inequality for three-state
  systems}, \href{https://doi.org/10.1103/PhysRevA.65.032118}{\emph{Phys. Rev.
  A} {\bfseries 65} (2002) 032118}.

\bibitem{PhysRevLett.88.040404}
D.~Collins, N.~Gisin, N.~Linden, S.~Massar and S.~Popescu, \emph{Bell
  inequalities for arbitrarily high-dimensional systems},
  \href{https://doi.org/10.1103/PhysRevLett.88.040404}{\emph{Phys. Rev. Lett.}
  {\bfseries 88} (2002) 040404}.

\bibitem{PhysRevLett.68.3259}
S.L.~Braunstein, A.~Mann and M.~Revzen, \emph{Maximal violation of bell
  inequalities for mixed states},
  \href{https://doi.org/10.1103/PhysRevLett.68.3259}{\emph{Phys. Rev. Lett.}
  {\bfseries 68} (1992) 3259}.

\bibitem{Acin:2002zz}
A.~Acin, T.~Durt, N.~Gisin and J.I.~Latorre, \emph{{Quantum nonlocality in two
  three-level systems}},
  \href{https://doi.org/10.1103/PhysRevA.65.052325}{\emph{Phys. Rev. A}
  {\bfseries 65} (2002) 052325}
  [\href{https://arxiv.org/abs/quant-ph/0111143}{{\ttfamily
  quant-ph/0111143}}].

\bibitem{Bernreuther:1997gs}
W.~Bernreuther, M.~Flesch and P.~Haberl, \emph{{Signatures of Higgs bosons in
  the top quark decay channel at hadron colliders}},
  \href{https://doi.org/10.1103/PhysRevD.58.114031}{\emph{Phys. Rev. D}
  {\bfseries 58} (1998) 114031}
  [\href{https://arxiv.org/abs/hep-ph/9709284}{{\ttfamily hep-ph/9709284}}].

\bibitem{Ethier:2021bye}
{\scshape SMEFiT} collaboration, \emph{{Combined SMEFT interpretation of Higgs,
  diboson, and top quark data from the LHC}},
  \href{https://doi.org/10.1007/JHEP11(2021)089}{\emph{JHEP} {\bfseries 11}
  (2021) 089} [\href{https://arxiv.org/abs/2105.00006}{{\ttfamily
  2105.00006}}].

\bibitem{Brivio_2017}
I.~Brivio and M.~Trott, \emph{Scheming in the {SMEFT}. . . and a
  reparameterization invariance!},
  \href{https://doi.org/10.1007/jhep07(2017)148}{\emph{Journal of High Energy
  Physics} {\bfseries 2017} (2017) }.

\bibitem{Grzadkowski:2010es}
B.~Grzadkowski, M.~Iskrzynski, M.~Misiak and J.~Rosiek, \emph{{Dimension-Six
  Terms in the Standard Model Lagrangian}},
  \href{https://doi.org/10.1007/JHEP10(2010)085}{\emph{JHEP} {\bfseries 10}
  (2010) 085} [\href{https://arxiv.org/abs/1008.4884}{{\ttfamily 1008.4884}}].

\bibitem{Aguilar-Saavedra:2018ksv}
D.~Barducci et~al., \emph{{Interpreting top-quark LHC measurements in the
  standard-model effective field theory}},
  \href{https://arxiv.org/abs/1802.07237}{{\ttfamily 1802.07237}}.

\bibitem{Romao:2016ien}
J.C.~Rom\~ao, \emph{{The need for the Higgs boson in the Standard Model}},
  \href{https://arxiv.org/abs/1603.04251}{{\ttfamily 1603.04251}}.

\bibitem{Falkowski:2016cxu}
A.~Falkowski, M.~Gonzalez-Alonso, A.~Greljo, D.~Marzocca and M.~Son,
  \emph{{Anomalous Triple Gauge Couplings in the Effective Field Theory
  Approach at the LHC}},
  \href{https://doi.org/10.1007/JHEP02(2017)115}{\emph{JHEP} {\bfseries 02}
  (2017) 115} [\href{https://arxiv.org/abs/1609.06312}{{\ttfamily
  1609.06312}}].

\bibitem{Helset:2017mlf}
A.~Helset and M.~Trott, \emph{{On interference and non-interference in the
  SMEFT}}, \href{https://doi.org/10.1007/JHEP04(2018)038}{\emph{JHEP}
  {\bfseries 04} (2018) 038}
  [\href{https://arxiv.org/abs/1711.07954}{{\ttfamily 1711.07954}}].

\bibitem{Azatov:2016sqh}
A.~Azatov, R.~Contino, C.S.~Machado and F.~Riva, \emph{{Helicity selection
  rules and noninterference for BSM amplitudes}},
  \href{https://doi.org/10.1103/PhysRevD.95.065014}{\emph{Phys. Rev. D}
  {\bfseries 95} (2017) 065014}
  [\href{https://arxiv.org/abs/1607.05236}{{\ttfamily 1607.05236}}].

\bibitem{Panico:2017frx}
G.~Panico, F.~Riva and A.~Wulzer, \emph{{Diboson interference resurrection}},
  \href{https://doi.org/10.1016/j.physletb.2017.11.068}{\emph{Phys. Lett. B}
  {\bfseries 776} (2018) 473}
  [\href{https://arxiv.org/abs/1708.07823}{{\ttfamily 1708.07823}}].

\bibitem{Franceschini:2017xkh}
R.~Franceschini, G.~Panico, A.~Pomarol, F.~Riva and A.~Wulzer,
  \emph{{Electroweak Precision Tests in High-Energy Diboson Processes}},
  \href{https://doi.org/10.1007/JHEP02(2018)111}{\emph{JHEP} {\bfseries 02}
  (2018) 111} [\href{https://arxiv.org/abs/1712.01310}{{\ttfamily
  1712.01310}}].

\bibitem{Azatov:2017kzw}
A.~Azatov, J.~Elias-Miro, Y.~Reyimuaji and E.~Venturini, \emph{{Novel
  measurements of anomalous triple gauge couplings for the LHC}},
  \href{https://doi.org/10.1007/JHEP10(2017)027}{\emph{JHEP} {\bfseries 10}
  (2017) 027} [\href{https://arxiv.org/abs/1707.08060}{{\ttfamily
  1707.08060}}].

\bibitem{Azatov:2019xxn}
A.~Azatov, D.~Barducci and E.~Venturini, \emph{{Precision diboson measurements
  at hadron colliders}},
  \href{https://doi.org/10.1007/JHEP04(2019)075}{\emph{JHEP} {\bfseries 04}
  (2019) 075} [\href{https://arxiv.org/abs/1901.04821}{{\ttfamily
  1901.04821}}].

\bibitem{Aoude:2019cmc}
R.~Aoude and W.~Shepherd, \emph{{Jet Substructure Measurements of Interference
  in Non-Interfering SMEFT Effects}},
  \href{https://doi.org/10.1007/JHEP08(2019)009}{\emph{JHEP} {\bfseries 08}
  (2019) 009} [\href{https://arxiv.org/abs/1902.11262}{{\ttfamily
  1902.11262}}].

\bibitem{Degrande:2021zpv}
C.~Degrande and J.~Touch\`eque, \emph{{A reduced basis for CP violation in
  SMEFT at colliders and its application to diboson production}},
  \href{https://doi.org/10.1007/JHEP04(2022)032}{\emph{JHEP} {\bfseries 04}
  (2022) 032} [\href{https://arxiv.org/abs/2110.02993}{{\ttfamily
  2110.02993}}].

\bibitem{Ball_2022}
R.D.~Ball, S.~Carrazza, J.~Cruz-Martinez, L.D.~Debbio, S.~Forte, T.~Giani
  et~al., \emph{The path to proton structure at 1{\%} accuracy},
  \href{https://doi.org/10.1140/epjc/s10052-022-10328-7}{\emph{The European
  Physical Journal C} {\bfseries 82} (2022) }.

\bibitem{Darme:2023jdn}
L.~Darm\'e et~al., \emph{{UFO 2.0 -- The Universal Feynman Output format}},
  \href{https://arxiv.org/abs/2304.09883}{{\ttfamily 2304.09883}}.

\bibitem{Alloul_2014}
A.~Alloul, N.D.~Christensen, C.~Degrande, C.~Duhr and B.~Fuks,
  \emph{{FeynRules} ~2.0~{\textemdash} a complete toolbox for tree-level
  phenomenology},
  \href{https://doi.org/10.1016/j.cpc.2014.04.012}{\emph{Computer Physics
  Communications} {\bfseries 185} (2014) 2250}.

\bibitem{Hahn_2001}
T.~Hahn, \emph{Generating feynman diagrams and amplitudes with {FeynArts} 3},
  \href{https://doi.org/10.1016/s0010-4655(01)00290-9}{\emph{Computer Physics
  Communications} {\bfseries 140} (2001) 418}.

\bibitem{Shtabovenko:2020gxv}
V.~Shtabovenko, R.~Mertig and F.~Orellana, \emph{{FeynCalc 9.3: New features
  and improvements}},
  \href{https://doi.org/10.1016/j.cpc.2020.107478}{\emph{Comput. Phys. Commun.}
  {\bfseries 256} (2020) 107478}
  [\href{https://arxiv.org/abs/2001.04407}{{\ttfamily 2001.04407}}].

\bibitem{Alwall:2014hca}
J.~Alwall, R.~Frederix, S.~Frixione, V.~Hirschi, F.~Maltoni, O.~Mattelaer
  et~al., \emph{{The automated computation of tree-level and next-to-leading
  order differential cross sections, and their matching to parton shower
  simulations}}, \href{https://doi.org/10.1007/JHEP07(2014)079}{\emph{JHEP}
  {\bfseries 07} (2014) 079} [\href{https://arxiv.org/abs/1405.0301}{{\ttfamily
  1405.0301}}].

\bibitem{Degrande:2020evl}
C.~Degrande, G.~Durieux, F.~Maltoni, K.~Mimasu, E.~Vryonidou and C.~Zhang,
  \emph{{Automated one-loop computations in the standard model effective field
  theory}}, \href{https://doi.org/10.1103/PhysRevD.103.096024}{\emph{Phys. Rev.
  D} {\bfseries 103} (2021) 096024}
  [\href{https://arxiv.org/abs/2008.11743}{{\ttfamily 2008.11743}}].

\bibitem{Costantini:2020stv}
A.~Costantini, F.~De~Lillo, F.~Maltoni, L.~Mantani, O.~Mattelaer, R.~Ruiz
  et~al., \emph{{Vector boson fusion at multi-TeV muon colliders}},
  \href{https://doi.org/10.1007/JHEP09(2020)080}{\emph{JHEP} {\bfseries 09}
  (2020) 080} [\href{https://arxiv.org/abs/2005.10289}{{\ttfamily
  2005.10289}}].

\bibitem{Al_Ali_2022}
H.A.~Ali, N.~Arkani-Hamed, I.~Banta, S.~Benevedes, D.~Buttazzo, T.~Cai et~al.,
  \emph{The muon smasher's guide},
  \href{https://doi.org/10.1088/1361-6633/ac6678}{\emph{Reports on Progress in
  Physics} {\bfseries 85} (2022) 084201}.

\bibitem{Accettura:2023ked}
C.~Accettura et~al., \emph{{Towards a Muon Collider}},
  \href{https://arxiv.org/abs/2303.08533}{{\ttfamily 2303.08533}}.

\bibitem{Aime:2022flm}
C.~Aime et~al., \emph{{Muon Collider Physics Summary}},
  \href{https://arxiv.org/abs/2203.07256}{{\ttfamily 2203.07256}}.

\bibitem{MuonCollider:2022xlm}
{\scshape Muon Collider} collaboration, \emph{{The physics case of a 3 TeV muon
  collider stage}},  \href{https://arxiv.org/abs/2203.07261}{{\ttfamily
  2203.07261}}.

\bibitem{Chiesa:2020awd}
M.~Chiesa, F.~Maltoni, L.~Mantani, B.~Mele, F.~Piccinini and X.~Zhao,
  \emph{{Measuring the quartic Higgs self-coupling at a multi-TeV muon
  collider}}, \href{https://doi.org/10.1007/JHEP09(2020)098}{\emph{JHEP}
  {\bfseries 09} (2020) 098}
  [\href{https://arxiv.org/abs/2003.13628}{{\ttfamily 2003.13628}}].

\bibitem{Ellis:2020unq}
J.~Ellis, M.~Madigan, K.~Mimasu, V.~Sanz and T.~You, \emph{{Top, Higgs, Diboson
  and Electroweak Fit to the Standard Model Effective Field Theory}},
  \href{https://doi.org/10.1007/JHEP04(2021)279}{\emph{JHEP} {\bfseries 04}
  (2021) 279} [\href{https://arxiv.org/abs/2012.02779}{{\ttfamily
  2012.02779}}].

\end{thebibliography}\endgroup

\end{document}